\documentclass[11pt]{article}
\usepackage{amsmath}
\usepackage{graphicx}
\usepackage{enumerate}
\usepackage{natbib}
\usepackage{url} 

\usepackage{array}
\usepackage{verbatim}

\usepackage[latin9]{inputenc}
\usepackage{amsthm}
\usepackage{amsmath}
\usepackage{amssymb}
\usepackage{setspace}
\usepackage{float} 
\usepackage{esint}

\usepackage{csvsimple}
\usepackage{hyperref}

\usepackage{authblk}

\theoremstyle{plain}

\theoremstyle{definition}

\usepackage{ifpdf}
\usepackage{longtable}
\usepackage{mathtools}

\usepackage{babel}
\providecommand{\definitionname}{Definition}
\providecommand{\theoremname}{Theorem}

\newcommand{\blind}{1}

\begin{document}

\def\spacingset#1{\renewcommand{\baselinestretch}%
{#1}\small\normalsize} \spacingset{1}


\if1\blind
{
  	\title{\bf A Riemann Manifold Model Framework for Longitudinal Changes in Physical Activity Patterns}
	
	\author[1]{Jingjing Zou \thanks{j2zou@ucsd.edu.
    DDS, SH, CLR, MJ  and LN were partially supported by funding from the National Cancer Institute (U54 CA155435-01);
    JB, AC, MJ, JZ, and LN were partially supported by funding from the National Institute of Diabetes and Digestive and Kidney Disease (R01DK114945); 
    JB, AC, CD, DDS, SH, and LN were partially supported by funding from the National Institute of Aging (PO1AG052352)} }

	\author[1]{Tuo Lin}
	
	\author[2]{Chongzhi Di}
	
	\author[1]{John Bellettiere}
	
	\author[3]{Marta M. Jankowska}
	
	\author[1,6]{Sheri J. Hartman}
	
	\author[4,5,6]{Dorothy D. Sears}
	
	\author[1]{Andrea Z. LaCroix}
	
	\author[5]{Cheryl L. Rock}
	
	\author[1,6]{Loki Natarajan }

	\affil[1]{Herbert Wertheim School of Public Health and Human Longevity Science, University of California, San Diego}
	
	\affil[2]{Division of Public Health Sciences, Fred Hutchinson Cancer Research Center}

    \affil[3]{Department of Population Sciences, Beckman Research Institute, City of Hope}

	\affil[4]{College of Health Solutions, Arizona State University; Department of Family Medicine, University of California, San Diego}
	
	\affil[5]{Department of Family Medicine, School of Medicine, University of California, San Diego}

	\affil[6]{UC San Diego Moores Cancer Center}
	
  \maketitle
} \fi

\if0\blind
{
  \bigskip
  \bigskip
  \bigskip
  \begin{center}
    {\LARGE\bf A Riemann Manifold Model Framework for Longitudinal Changes in Physical Activity Patterns}
\end{center}
  \medskip
} \fi


\begin{abstract}
Physical activity (PA) is significantly associated with many health outcomes.
The wide usage of wearable accelerometer-based activity trackers in recent years has provided a unique opportunity for in-depth research on PA and its relations with health outcomes and interventions.
Past analysis of activity tracker data relies heavily on aggregating minute-level PA records into day-level summary statistics, in which important information of PA temporal/diurnal patterns is lost. 
In this paper we propose a novel functional data analysis approach based on Riemann manifolds for modeling PA and its longitudinal changes. 
We model smoothed minute-level PA of a day as one-dimensional Riemann manifolds and longitudinal changes in PA in different visits as deformations between manifolds.
The variability in changes of PA among a cohort of subjects is characterized via variability in the deformation. 
Functional principal component analysis is further adopted to model the deformations and PC scores are used as a proxy in modeling the relation between changes in PA and health outcomes and/or interventions.
We conduct comprehensive analyses on data from two clinical trials: Reach for Health (RfH) and Metabolism, Exercise and Nutrition at UCSD (MENU),
focusing on the effect of interventions on longitudinal changes in PA patterns and how different modes of changes in PA influence weight loss, respectively. 
The proposed approach reveals unique modes of changes including overall enhanced PA, boosted morning PA, and shifts of active hours specific to each study cohort.
The results bring new insights into the study of longitudinal changes in PA and health and have the potential to facilitate designing of effective health interventions and guidelines.

\end{abstract}

\noindent%

\spacingset{1.9} 

\section{Introduction}

Physical activity (PA) has been linked to health in many epidemiological studies. Physical inactivity and sedentary behavior are known risk factors for cardiovascular disease, cancer and all-cause mortality ( \cite{LaMonte:2018jb, LaCroix:2019eb, Chastin:2019bw, Bellettiere2019b, Bellettiere2019a, Bellettiere:2020ie, Parada:2020fi, Ramakrishnan2021, Glass2021, Walker2021}). 

Activity trackers based on sensor devices such as accelerometers enables detailed tracking of daily PA at minute-level in free-living environment. 
In studies using activity trackers, participating subjects wear the device in body parts such as hip, thigh or wrist, and perform their usual daily activities non-intrusively (\cite{Prince:2008kq,Colley:2011we,Dyrstad:2014ik,Adamo:2009en,Ainsworth:2014cr}).
For systematic reviews and meta-analyses on data from accelerometer-based trackers, see \cite{Migueles:2017en, Stamatakis:2019kj, Ekelund:2019dl}.

The emerging usage of activity trackers in studies provides an unprecedented opportunity for in-depth research on physical activity and sedentary behavior. 
In particular, recording PA of subjects with activity trackers in multiple visit periods, including the baseline and follow-ups, not only reveals information on PA during each period, but also enables estimation of longitudinal changes in PA between periods/visits. 
As a result, the impact of interventions, if incorporated in the study, can be evaluated by examining the association between the intervention and longitudinal changes in PA. 
Moreover, the effect of longitudinal changes in PA on health outcomes such as obesity and physiological measurements relevant to diseases can be studied in correlative analysis of changes in PA and outcomes of interest. 

The aforementioned analyses require statistical modeling of PA with minute-level activity tracker data and of the longitudinal changes in PA. 
Past studies of activity tracker data rely heavily on summarizing minute-level data into single day-level measurements such as assessed sedentary time (SED), light-intensity physical activity (LIPA), and moderate-to-vigorous physical activity (MVPA) (\cite{Nader:2008hm,Matthews:2002ju,Loprinzi:2016jx,Fuzeki:2017eb}).
Each of the day-level measurements summarizes a facet of the activity tracker data. However, important information can be lost in aggregating minute-level PA records. 
In particular, the temporal dependence in PA at different times of a day cannot be assessed after day-level reductions, which prevents the inclusion of PA diurnal patterns in the scope of further studies.

Functional data analysis (FDA) approaches that preserve the original format of the minute-level data instead of aggregating them into summary statistics have been adopted in explaining variability in PA, especially in temporal activity patterns, among subjects within one cohort. 
In these studies, PA is usually characterized as a subject-specific $X_{ij}(t)$ indexed by chronological time $t$, where $i$ and $j$ index subjects and visits respectively. Variability in PA is often modeled with functional principal component analysis (fPCA) type of variance decomposition method.
For example, 
\cite{Li:2014fo} modeled jointly the energy expenditure and interruptions to sedentary behaviors with fPCA.
\cite{Goldsmith:2015cc} proposed a generalized multilevel fPCA model to study minute-level PA recorded by activity trackers of subjects in multiple days. 
\cite{Li:2015hg} extended the model in \cite{Goldsmith:2015cc} and proposed a three-level functional data model allowing daily records to be nested in weeks. 
For details of multilevel fPCA and its applications to longitudinal functional data, see, for example, \cite{Di:2009jq} and \cite{Greven:2010ik, Greven:2011jg}.

Studies focusing on other aspects of PA records and functional data in general have also been conducted. 
\cite{Xiao:2015jf} proposed a covariate-dependent functional model to quantify the lifetime circadian rhythm of PA. 
\cite{Shou:2014iz} discussed a structured fPCA to handle multiple levels of variation generated by nested and crossed study designs.
\cite{Goldsmith:2016go} adopted function-on-scalar regression methods to relate 24-hour diurnal PA patterns with covariates.
\cite{Xu:2019bd} implemented an fPCA mixed model to study multiple daily PA records and examined the association between activity patterns and health outcomes. 
\cite{Wrobel2019} introduced an fPCA approach for estimating a template for aligning functions in temporal domain and a time warping method for registration of subject-specific functions to the template. 
\cite{choi2018temporal} employed temporal registration procedures prior to classifications of activity types using machine learning methods.
\cite{srivastava2011registration} proposed a geometric framework for separating the phase and the amplitude variability in functional data. The registration is based on warping using the Fisher-Rao Metric.
\cite{anirudh2016elastic} approached the problem of functional data registration based on transported square-root velocity fields.
\cite{kurtek2017geometric} developed a Bayesian model for pairwise nonlinear registration of functional data and explored statistical inference tools including $k$-means for clustering.
\cite{Reuter:2020cd} proposed a two-stage clustering method for sedentary behavior (SB) and examined associations between SB patterns and longitudinal physical functioning with a mixed model.



Despite the recent development of FDA approaches for minute-level PA data, it remains unclear how to directly model longitudinal changes in PA diurnal patterns and how to correlate changes in PA with health interventions and outcomes.
Motivated by two clinical trials, RfH and MENU,
we propose a new framework by applying Riemann manifold theory to the analysis of longitudinal PA records. 
The model framework consists of multiple stages. 
First, pre-processed and smoothed minute-level PA records are characterized as one-dimensional (1D) Riemann manifolds (curves).
Next, the longitudinal changes in PA are modeled as deformations between the manifolds/curves. 
Specifically, the deformations are modeled by diffeomorphisms governed by elements in a reproducing kernel Hilbert space that satisfy minimal-energy constraints (\cite{Beg:2005gr, Vaillant:2004fra, Charlier:2015fha}).
Results on diffeomorphisms have been used in matching and registration of medical images such as structural brain fMRI, see, for example, \cite{Hernandez:2009dp}.
However, to our best knowledge, this is one of the first studies of longitudinal changes in physical activity utilizing the power of diffeomorphism theory.
We further model the variability in longitudinal changes of PA with fPCA. 
Principal components (PC) that are capable of explaining a majority of the variability are examined as the main modes of changes in PA patterns, and the corresponding projection coefficients (scores) are used to characterize the composition of different modes of variation in subject-specific PA changes.
Finally, relations between longitudinal changes in PA diurnal patterns and health outcomes and/or interventions are analyzed via correlative studies of PC projection coefficients.

An important advantage of the proposed approach is that it models not only changes in PA magnitude at each fixed time point of a day in different visit periods, but also temporal/phase shifts of activity patterns. 
The capability of capturing phase changes is critical in the study of PA, as subjects can experience change of circadian rhythms and may exhibit a shift of active hours including changes in sleep/wake up cycles associated with health outcomes and interventions (\cite{Montaruli:2017ji, Khan:2021ei}).
The proposed approach enables identification of such important changes in PA patterns and results can facilitate the 
design of effective interventions to promote beneficial PA habits.


The paper is organized as follows. 
Section \ref{sect: data motivation} introduces the two clinical trials that motivate the proposed approach.
Section \ref{sect: model and method} describes the Riemann manifold model framework. 
Section \ref{sect: estimation methods} delineates the estimation pipeline for parameters in the proposed model. 
Section \ref{sect: simulation} conducts simulation studies to evaluate the proposed approach.
Sections \ref{sect: data analysis 1} and \ref{sect: data analysis 2} conduct comprehensive data analyses on data from the RfH and MENU studies.
Technical details and additional simulation and data analysis results are available in the supplementary material.

\section{Data Motivation} \label{sect: data motivation}

The model and method proposed in this paper are motivated by two clinical trials: Reach for Health (RfH) and Metabolism, Exercise and Nutrition at UCSD (MENU).
Both studies were part of the NIH-funded Transdisciplinary Research on Energetics and Cancer (TREC) Study at University of California, San Diego (UCSD) from 2011 to 2017. 
The RfH study is a six-month clinical trial that involved 333 overweight, postmenopausal early-stage breast cancer survivors. 
Each of the subjects was randomly assigned to either metformin treatment or placebo, combined with either a lifestyle-based intervention or placebo. 
The primary aim was to test the effect of treatments on health outcomes, especially weight loss as measured by the change in BMI (\cite{Patterson:2018ck}).
The MENU study is a 12-month behavioral intervention study on 245 overweight and otherwise healthy women. 
Each subject was randomly assigned to one of the three diet treatment groups: 
a lower fat (20\% of energy) and higher carbohydrate (65\% of energy) diet; a lower carbohydrate (45\% energy) and higher monounsaturated fat (35\% energy) diet; or a walnut-rich (35\% fat) and lower carbohydrate (45\%) diet (\cite{Rock:2016ko, Le:2016ft, Patterson:2016ii}). 
The goal was to study the role of dietary macronutrient composition on weight loss. 

In both studies, multi-day PA at baseline and follow-up periods were measured by the GT3X ActiGraph (ActiGraph, LLC; Pensacola, FL; \url{www.actigraphcorp.com}), a research-grade activity tracker with built-in triaxial accelerometer. 
Activity data were collected at high-resolution of 30 Hz, then processed into per minute PA vector magnitudes (VM, the $L^2$ norm of the triaxial PA records) (\cite{Bassett:2012jt, Xu:2019bd}).
Our goal is to characterize longitudinal changes in PA diurnal patterns and examine the relations between changes in PA patterns and health outcome (for the MENU study) and interventions (for the RfH study).

\section{The Model Framework} \label{sect: model and method}


\subsection{A Riemann Manifold Model for Physical Activity} \label{subsect: manifold}


Suppose there are $N$ subjects in the study.
For the $i$th subject, daily PA is recorded by the activity tracker in periods $k = 0,\dots,K$, where $k = 0$ denotes the baseline visit and $k = 1,\dots, K$ denote the follow-ups.
The number of visits is assumed to be the same for all subjects.
Data of each period consist of daily minute-level PA VM records of several consecutive days.
The multi-day PA records in each period are averaged at each of the chronological time $t$ of a day and smoothed to reduce noise (details in Section \ref{sect: estimation methods}). 
For the $i$th subject, 
the averaged and smoothed PA measured at the $k$th period in the longitudinal study is modeled as a one-dimensional Riemann manifold/curve $X_i^{(k)}$ with image on $[0,T] \times [0, M] \subset \mathbb{R}^2$, where
$t \in [0, T]$ is the index for chronological time of a day and 
$M$ denotes the maximal admissible PA magnitude for all subjects.

For subject $i$, we characterize the longitudinal change in the PA from visit $(k-1)$ to $k$ as
\begin{equation}\label{eq: deformation of manifolds}
	X_i^{(k)} = \phi(\boldsymbol{v}_i^{(k)}, \cdot) \circ X_i^{(k-1)},
\end{equation}
where $\phi(\boldsymbol{v}_i^{(k)}, \cdot) $ is a subject-specific diffeomorphism that maps the PA curve $X_i^{(k-1)}$ at the $(k-1)$th period to the PA curve $X_i^{(k)}$ at the $k$th period.
The symbol $\circ$ denotes the composite of two functions.
The diffeomorphism is assumed to be governed by $\boldsymbol{v}_i^{(k)} $, which is a subject-specific element in $L^2([0,1], V)$.
The space $V$ contains vector fields defined on $[0,T] \times [0, M]$, 
and each element $ \boldsymbol{v}\in L^2([0,1], V)$ is a set/series of time-varying vector fields indexed by $\tau \in [0,1]$ with finite $L^2$ norm: $\int_0^1 \|v_\tau\|^2 d\tau < \infty$, where $v_\tau$ denotes the value of $\boldsymbol{v}$ at a fixed index $\tau$.
Note $\tau$ should not be mixed up with the chronological time $t$ in a day.

The diffeomorphism $\phi(\boldsymbol{v}_i^{(k)}, \cdot)$ in \eqref{eq: deformation of manifolds} is obtained by \textit{flowing} $\boldsymbol{v}_i^{(k)} $.
That is, for any $\boldsymbol{v} \in L^2([0,1], V)$,
let $\psi_{\boldsymbol{v}}$ be the solution to the ordinary differential equations 
$\frac{\partial}{\partial \tau} \psi_{\boldsymbol{v}}(\tau, \cdot) = v_\tau \circ \psi_{\boldsymbol{v}}(\tau, \cdot)$ with the initial condition that $\psi_{\boldsymbol{v}}(0, \cdot) $ is the identity mapping.
The deformation $\phi(\boldsymbol{v}, \cdot) = \psi_{\boldsymbol{v}}(1, \cdot) $ is defined to be the final state of $\psi_{\boldsymbol{v}}$.
Intuitively, $\boldsymbol{v}$ can be considered as a series of forces that ``drag'' the PA curve from one period to another at each $\tau \in [0,1]$. 
At each step $\tau $, the force drags the current PA curve by the amount of $v_\tau$ until the final step $\tau = 1$. 
Equation \eqref{eq: deformation of manifolds} states that
with the subject-specific vector field $\boldsymbol{v}_i^{(k)}$, 
the total deformation applied to the PA curve at period $(k-1)$ is $\phi(\boldsymbol{v}_i^{(k)}, \cdot) = \psi_{\boldsymbol{v}_i^{(k)}}(1, \cdot)$ to obtain the PA curve at period $k$.
For technical details on the diffeomorphisms, see \cite{Charlier:2015fha} and the references therein.

The $\boldsymbol{v}_i^{(k)}$ at each $\tau \in [0,1]$ is further assumed to follow a reproducing kernel Hilbert space (RKHS) representation: 
\begin{equation} \label{eq: rkhs}
	v_\tau(\cdot) = \sum_{j=1}^{n_g} K_{V}(\psi_{\boldsymbol{v}}(\tau, c_j), \cdot)  \, m_{j,\tau},
\end{equation}
where $\{c_j: j = 1,\dots, n_g\} \subset \mathbb{R}^2$ is a set of pre-selected control points of total number $n_g$. 
$K_{V}(x, y) = \exp(-\|x-y\|^2 / (2 \sigma^2_{V})) \mathbf{I}_2 $ is the Gaussian isotropic kernel with a fixed rigidity parameter $\sigma^2_{V}$,
and $m_{j, \tau}\in \mathbb{R}^2$ is the momentum of the deformation at the $j$th control point at $\tau$. 
Intuitively, \eqref{eq: rkhs} states the force that drags the PA curve is determined by forces (momenta) on a finite number of points on the curve, for example, at a fixed number of minutes in a day. 
The dragging forces applied to non-control points can be represented as weighted averages of those on the control points.


Given $X_i^{(k-1)}$ and $X_i^{(k)}$,
the deformation $\phi(\boldsymbol{v}_i^{(k)}, \cdot)$ that satisfies \eqref{eq: deformation of manifolds} is not unique. 
Further constraints are needed to uniquely define the vector field $\boldsymbol{v}_i^{(k)}$ and its associated deformation $\phi(\boldsymbol{v}_i^{(k)}, \cdot)$. Here we adopt the constraint on the deformation energy 
$\int_0^1 \|v_\tau\|^2 d\tau$.
We model the change in the $i$th subject's PA between the $(k-1)$th and the $k$th periods as the deformation $\phi(\boldsymbol{v}_i^{(k)}, \cdot)$ that satisfies \eqref{eq: deformation of manifolds} with $\boldsymbol{v}_i^{(k)}$ minimizing the energy.
Subject to the constraint on the deformation energy,
the $\boldsymbol{v}_i^{(k)}$ that satisfies \eqref{eq: deformation of manifolds} 
is unique and is determined solely by the values of $\boldsymbol{v}_i^{(k)}$ at $\tau = 0$ (\cite{Charlier:2015fha}). 

With the above model, the variability in the longitudinal changes in PA among subjects in the cohort can be characterized by the subject-specific random vector fields $\{\boldsymbol{v}_i^{(k)}: i = 1, \cdots, N, k = 1, \cdots, K \}$.
Furthermore, with the RKHS representation \eqref{eq: rkhs}, the \textit{initial momenta} $ (m^i_{1, 0}, \cdots, m^i_{n_g, 0})$ (initial meaning at $\tau=0$) fully determine the deformation, and we denote the initial momenta by ${m}^i$ when there is no ambiguity.

\subsection{Functional Principal Component Analysis (fPCA) for the Deformations} \label{sect: fPCA}

The variability in longitudinal changes in PA among subjects in a cohort can be represented further by a function principal component model. 
For simplicity, we first assume $K = 1$ which means there is only one follow-up after the initial (baseline) visit/period. 
The case of multiple follow-ups can be readily extended:  
an extra dimension can be introduced for the visits, so that the domain of the vector fields at each $\tau$ becomes $\{1, \cdots, K\} \times E$, and for each $k = 1, \cdots, K$, the sub-vector-field $v: \{k\} \times E \rightarrow E$ denotes the vector field governing the deformation from the PA curve at the $(k-1)$th visit to the $k$th visit. Then the same analysis can be applied to the extended-dimension vector fields.

Consider the principal component representation of the initial momenta
$ m^i = \bar{m} + \sum_{l = 1}^{\infty} a_{il} \mu_l $,
where $\bar{m}$ is the mean of the momenta and $\{\mu_l: l = 1, \cdots, \infty\} \subset \mathbb{R}^{n_g \times 2}$ are the principal components (PC).
The $a_{il} = \langle m^i - \bar{m}, \mu_l \rangle$, where $\langle \cdot, \cdot \rangle$ denotes the Frobenius inner product on $ \mathbb{R}^{n_g \times 2}$, is the projection coefficient (score) of the mean-subtracted $ m^i $ on the $l$th PC.
Assuming the variability in the deformations of subjects in the cohort can be well explained by a finite number $N_{pc}$ of principal components, we can write approximately
$ m^i = \bar{m} + \sum_{l = 1}^{N_{pc}} a_{il} \mu_l $.
In practice, the value of $N_{pc}$ can be determined by researchers based on criteria including the proportion of variance explained by the top PCs.

\subsection{Correlative study with health outcomes} \label{sect: models for outcomes}

Two types of questions are often of interest to researchers: 1. how do PA diurnal patterns change as responding to interventions designed for encouraging more active lifestyle, and 2. how do longitudinal changes in PA affect health outcomes. To address these two questions, we model the relation between changes in PA and intervention/health outcomes with regressions, in which the longitudinal changes in PA are modeled via the PC projection coefficients/scores $\{a_{il}\}$. The model is
$ a_{il} = f_l(W_i, Z_i)	+ \epsilon_{il} $
for effect of intervention $W_i$ and covariates $Z_i$ on changes in PA.
Here the regression is carried out separately for each PC indexed by $l$.
The model is
$ Y_i = f(a_{i1}, \cdots, a_{i N_{pc}}, Z_i) + \epsilon_i $
for effects of changes in PA on health outcome(s) $Y$. 
In both models the $\epsilon$ are independent and identically distributed (iid) errors.

\section{Estimation Methods} \label{sect: estimation methods}

In this section we delineate the estimation procedure for model parameters in Section \ref{sect: model and method}. 
We focus on two groups of parameters: the parameters in modeling the longitudinal changes in PA via diffeomorphisms
and the parameters in representing the group-level variability of changes in PA with the fPCA model. 
Parameters in models in Section \ref{sect: models for outcomes} can be estimated with standard methods for regressions.



\subsection{Estimation of Diffeomorphisms and Initial Momenta} \label{sect: est momenta}

\begin{figure}[H]
	\centering
	\includegraphics[width=1\linewidth]{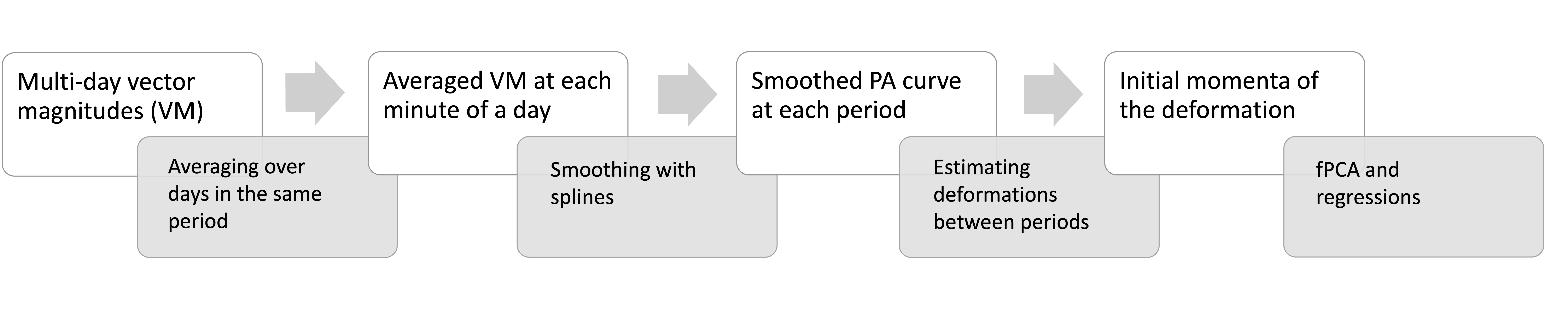}
	\caption{\small Steps in the estimation pipeline. }
	\label{fig: flow}
\end{figure}

In this section we focus on the estimation of the deformation vector fields $\{\boldsymbol{v}_i^{(k)}\}$ and its associated initial momenta $\{m^i\}$ in the diffeomorphism model described in Section \ref{subsect: manifold}. 
The flowchart in Figure \ref{fig: flow} summarizes the steps involved, and in what follows we introduce the details in each step.

In practice, each subject's PA measured by the accelerometers of activity trackers are available to the researcher in the discrete format. 
In each visit/period, there are usually several consecutive days of PA records. 
For each day, pre-processed PA records are available in every minute of a day.
For each subject, the multi-day PA records in the same period are further averaged to reduce noise. 

\begin{figure}[h]
	\centering
	\includegraphics[width=0.8\linewidth]{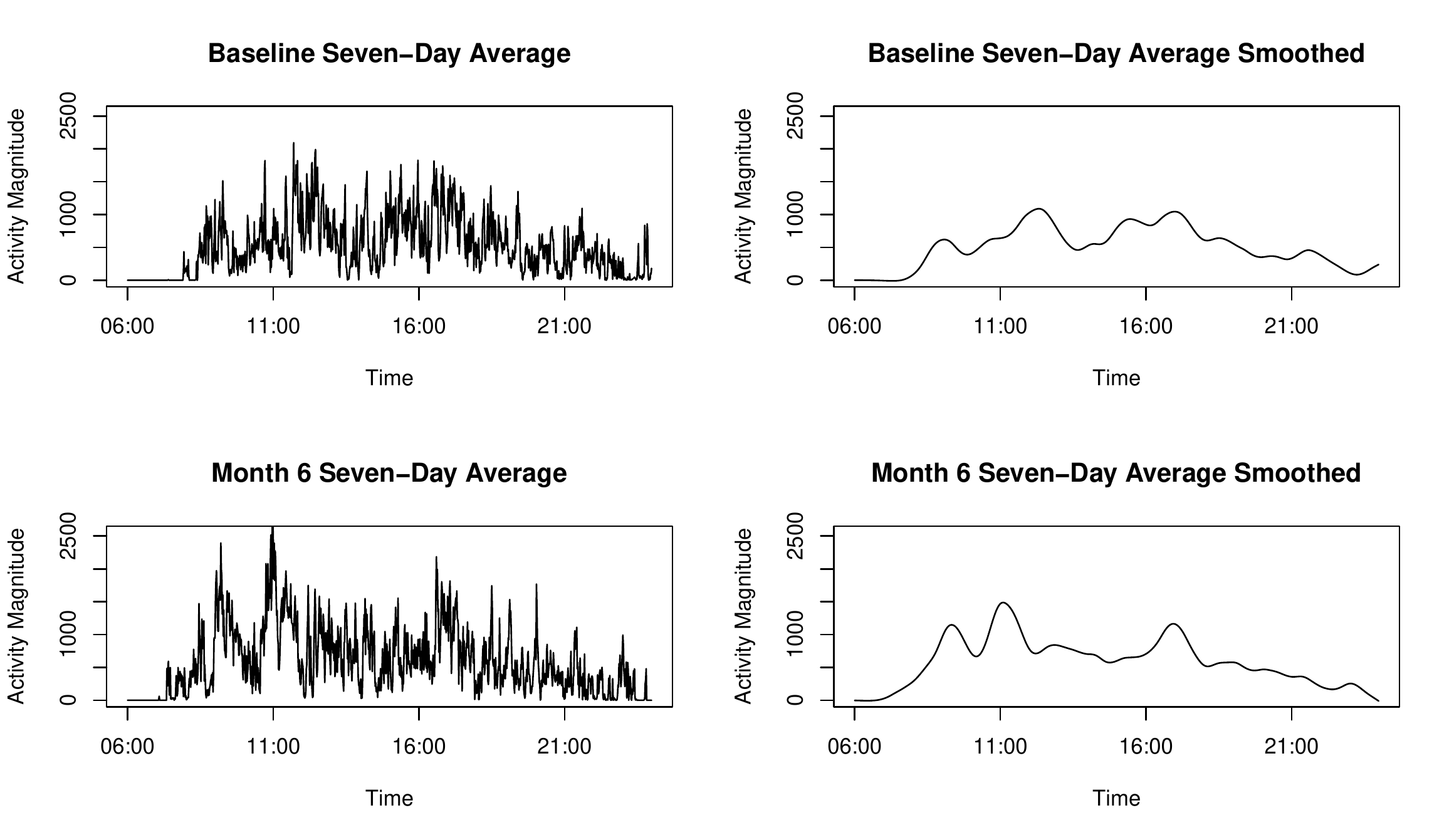}
	\caption{\small Example of PA curves at baseline and month-6 before and after smoothing for one subject in the RfH study.}
	\label{fig:curve_example}
\end{figure}

The averaged PA records in each visit are then smoothed using splines with function \textit{smooth.spline()} in software R 4.0.3 (\cite{Rsoftware}) to further reduce noise caused by minor fluctuations of the activity tracker that are irrelevant to meaningful movements.
We choose the degree of freedom (trace of smoothing matrix) $df = 25$ in spline smoothing based on exploratory analysis of datasets from both RfH and MENU studies so that the smoothed PA records preserve visible diurnal PA patterns while reducing noise induced by the trackers.
Note here we use the same degree of freedom in smoothing the RfH and MENU PA records, as both studies use the GT3X Actigraph activity tracker to measure subjects' PA, leading to comparable noise levels in PA records.
It is also important to use consistent smoothing and scaling parameters for all subjects' PA at all periods to preserve informative variability in the PA.
Figure \ref{fig:curve_example} displays the smoothed PA curves at baseline visiting period and month 6 from an example subject in the RfH study. 
A sensitivity analysis on the smoothing parameter is available in the supplementary material in which multiple values of the smoothing parameters are examined and options are discussed for practical use of the smoothing procedure.
In summary, the proposed approach is generally robust when the smoothing parameter is in a reasonable range. 
When smoothing is not adequate, the remaining noise can influence the results, especially estimated PCs that explain less variability of the data. 
For a different study using other activity tracker devices, it is recommended to conduct exploratory analysis and/or lab calibrations of the data to determine the appropriate smoothing parameter.

For each subject $i$ and period $k$, we estimate the subject-specific deformation between smoothed PA curves $X_i^{(k-1)}$ and $X_i^{(k)}$.
The estimation of the deformation that transforms a subject's PA curves from one period (the \textit{source}) to another (the \textit{target}) is formulated as an optimization problem: the goal is to find the deformation that minimizes a dissimilarity metric $g$ between the deformed source curve and the target curve, while constraining on the total energy $\int_0^1 \|v_\tau\|^2 d \tau$ of the deformation.
In computing the optimal deformation, the optimization problem with constraint on energy is transformed into an unconstrained problem with the following penalized objective function
\begin{equation} \label{eq: unconstrained}
	\frac{\gamma_W}{2} g(\tilde{X}_i^{(k)}, X_i^{(k)}) + \frac{\gamma_V}{2} \int_0^1 \|v_\tau\|^2 d \tau,
\end{equation}
where $g$ is the norm of the difference between the observed target curve $X_i^{(k)}$ and the estimated target curve $\tilde{X}_i^{(k)} =  \phi({\boldsymbol{v}}_i^{(k)}, \cdot) \circ X_i^{(k-1)}$ resulting from deforming $ X_i^{(k-1)}$ according to vector fields ${\boldsymbol{v}}_i^{(k)}$. 
The norm of the difference is defined based on Gaussian kernels between two discretized curves (for details see eq. (62) of \cite{Charlier:2015fha}).
Here $\gamma_V$ and $\gamma_W$ are pre-selected penalty parameters for the energy and dissimilarity terms.
The goal is to find the optimal subject-specific vector fields ${\boldsymbol{v}}_i^{(k)}$ for \eqref{eq: unconstrained}.

As discussed previously, the vector fields ${\boldsymbol{v}}_i^{(k)}$ are assumed to follow an RKHS representation \eqref{eq: rkhs} and are fully determined by the initial momenta $m^i$ at a set of pre-selected control points. 
Therefore, optimizing the objective \eqref{eq: unconstrained} over admissible values of ${\boldsymbol{v}}_i^{(k)}$ is equivalent to finding the optimal initial momenta for the objective function.
A natural choice of the control points is the set of available time points (minutes) and associated PA magnitudes in the data.
The optimization is conducted using the Hamiltonian method with the function ``fsmatch\_tan()'' in the package \textit{fshapesTK} (\cite{Charlier:2015fha}).
An example Matlab script to estimate subject-specific deformation is available in the supplementary material.

\begin{figure}[h]
	\centering
	\includegraphics[width=1\linewidth]{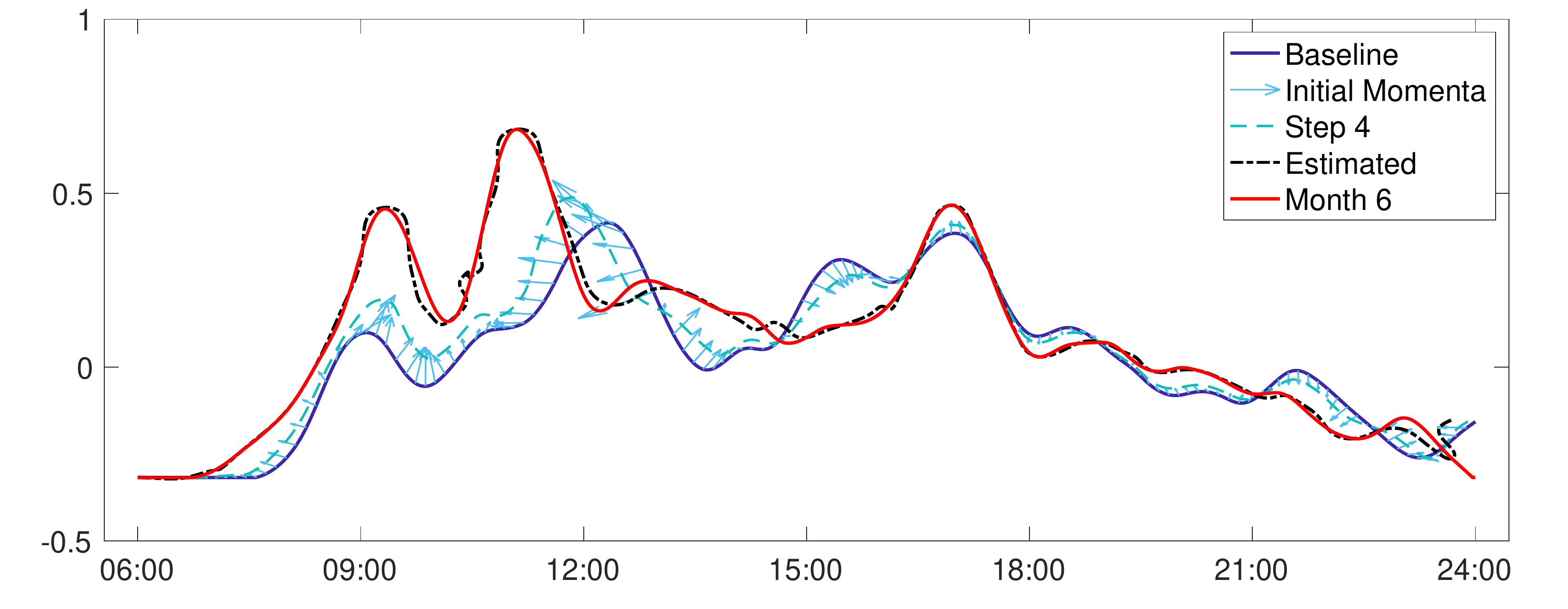}
	\caption{\small RfH study: estimated deformations of one example subject. Baseline PA curve (navy) with estimated initial momenta at each control point (light blue arrows on top of the baseline curve), middle step ($\tau = 4/11$) of deformation (dashed lines), and estimated (black dashed) and actual month-6 curves (red). $y$-axis: centered and scaled PA magnitude. $x$-axis: time of a day.} 
	\label{fig:deform_example}
\end{figure}

As an example,
Figure \ref{fig:deform_example} shows the estimated deformation and initial momenta of one subject in the RfH study.
Blue arrows illustrate the initial momenta at control points (arrows are drawn every 10 mins for a better visualization).
The initial momenta capture several key characteristics of changes in PA patterns.
PA levels increase from baseline to month 6 most significantly in time windows 6am-12pm and 1-3pm, reflected by the upward pointing arrows.
The arrows of momenta also signal two visible shifts of active hours: the small peak centered around 9am at baseline visit shifts slightly to the right after six months, while the peak centered around 12pm shifts significantly to the left.

Note that the second term in the objective function \eqref{eq: unconstrained} is a penalty on the deformation energy $ \int_0^1 \|v_\tau\|^2 d \tau $, which imposes a penalty on the initial momenta $m^i$ with equal weights on its $x$ and $y$ coordinates.
An alternative approach is to impose different weights on $x$ and $y$ coordinates. 
How to choose the weights, however, is a substantive question, determined by the emphasis on the temporal ($x$) and vertical ($y$) changes of the PA.
Alternatively, one can apply scaling to $x$ and $y$ coordinates of the PA curves to control the ranges of both coordinates. 
A smaller range results in less energy consumption and thus is equivalent to less penalty weight.
In our analysis for the RfH and MENU data, we adopt the latter approach and center and scale the time indices and PA magnitudes for them to be in comparable scales, so that the temporal shift and vertical changes of PA patterns are emphasized equally.

\subsection{fPCA of Momenta}

The estimated initial momenta for the $i$th subject $\hat{m}^i: = (\hat{m}^i_{1, 0}, \cdots, \hat{m}^i_{n_g, 0})$ is a vector field defined on the $n_g$ control points and stored in an $n_g \times 2$ matrix.
At the $l$th control point $\hat{m}^i_{l}$ is a vector in $\mathbb{R}^2$ indicating directions and magnitudes of the momenta in both the temporal ($x$-axis) and vertical directions ($y$-axis). The $y$ direction indicates changes in PA levels, while the $x$ direction indicates temporal shifts of PA patterns. 

Suppose a finite number of functional principal components (PC) are adequate in explaining the majority of the variability in the deformation from the baseline to the follow-up PA curves for all subjects. 
Calculating the PCs of dimension $\mathbb{R}^{n_g \times 2}$ under the Frobenius inner product is equivalent to calculating the PCs of vectors obtained from concatenating the $\hat{m}^i$ matrices.
Note the mean of the estimated initial momenta needs to be subtracted prior to estimating the PCs.
To interpret and visualize the PCs, we reshape the estimated initial momenta for each PC to the original dimensionality $n_g \times 2$, and ``flow'' (as described in Section \ref{subsect: manifold}) a pre-selected template curve with the deformation governed by the estimated initial momenta using the Matlab function ``shoot\_and\_flow\_tan()'' in package \textit{fshapesTK} (\cite{Charlier:2015fha}). 


\section{Simulation Studies} \label{sect: simulation}
\begin{figure}[h] 
	\centering
	\begin{minipage}[b]{0.8\textwidth}  
		\centering
		\includegraphics[width=\textwidth]{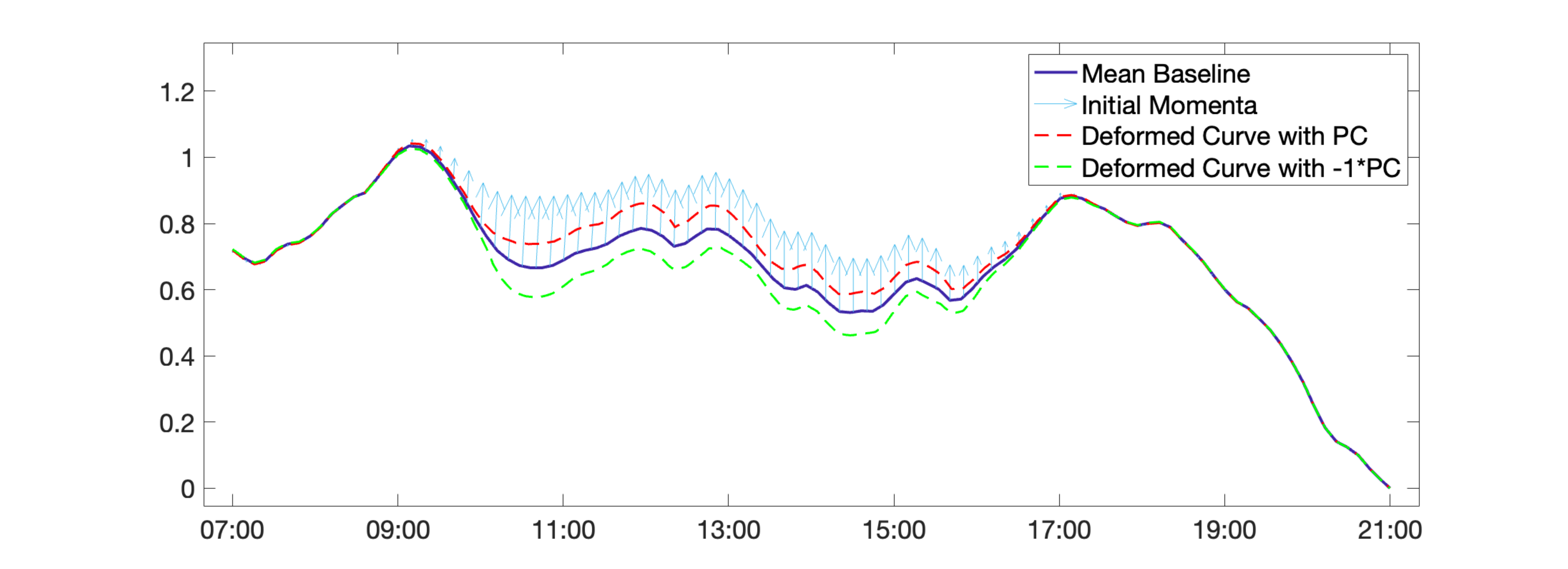}
	\end{minipage} 
	\begin{minipage}[b]{0.8\textwidth}  
		\centering
		\includegraphics[width=\textwidth]{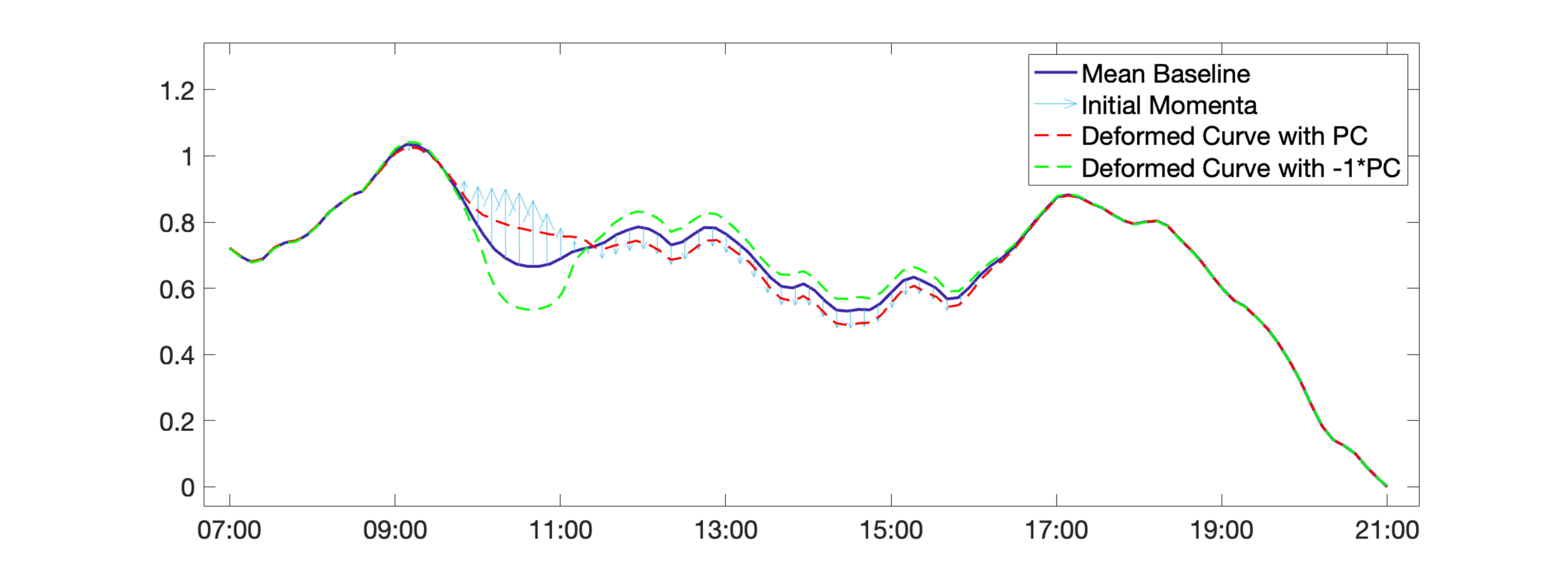}
	\end{minipage} 
	\begin{minipage}[b]{0.8\textwidth}  
		\centering
		\includegraphics[width=\textwidth]{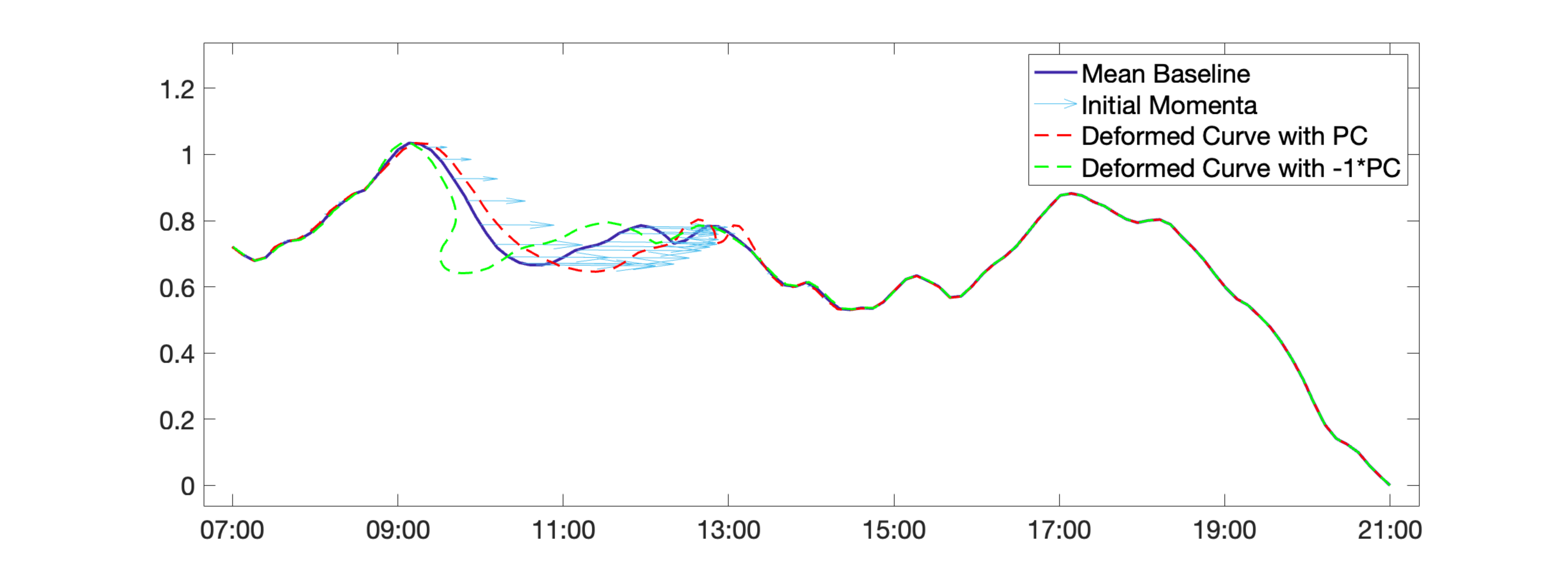}
	\end{minipage}
	\caption{From top to bottom: actual initial momenta (blue arrows) of PC 1, 2, and 3 and deformations when applied to the mean baseline curve (solid navy) to deform into the follow-up curves. Dashed red line is deformed curve with $1 \times$ PC and green is $-1 \times $ PC} \label{fg: PC_actual}
\end{figure}

In this section we evaluate the proposed approach with a synthetic dataset and compare the proposed approach to two existing methods. 
First we generate a smooth baseline curve using the averaged baseline PA of the MENU study data, which consists of PA vector magnitudes (VM) at each minute from 7am to 9pm. 
Here we use the MENU study data to generate the synthetic data in order to reflect the realistic PA patterns in the simulations, and the time window of 7am-9pm is selected for demonstrative purposes only.
Then we create $3$ principal components (PC) by first manually creating three sets of initial momenta and then running functional PCA (with function \textit{FPCA()} of R package \textit{fdapace} by \cite{fdapace}) to extract orthogonal eigen-functions of them. 
The resulting mutual-orthogonal initial momenta represent three modes of changes in PA diurnal patterns from the baseline to the follow-up visit. 
Figure \ref{fg: PC_actual} shows the initial momenta of the PCs as well as the follow-up curves resulting from applying the PCs on the mean baseline PA curve.
The first PC is a general increase of PA levels throughout the day, 
the second is a local boost of PA, and
the third is mainly a time shift of active hours to later times of the day.

\begin{figure}[h]
	\centering
	\includegraphics[width=0.85\linewidth]{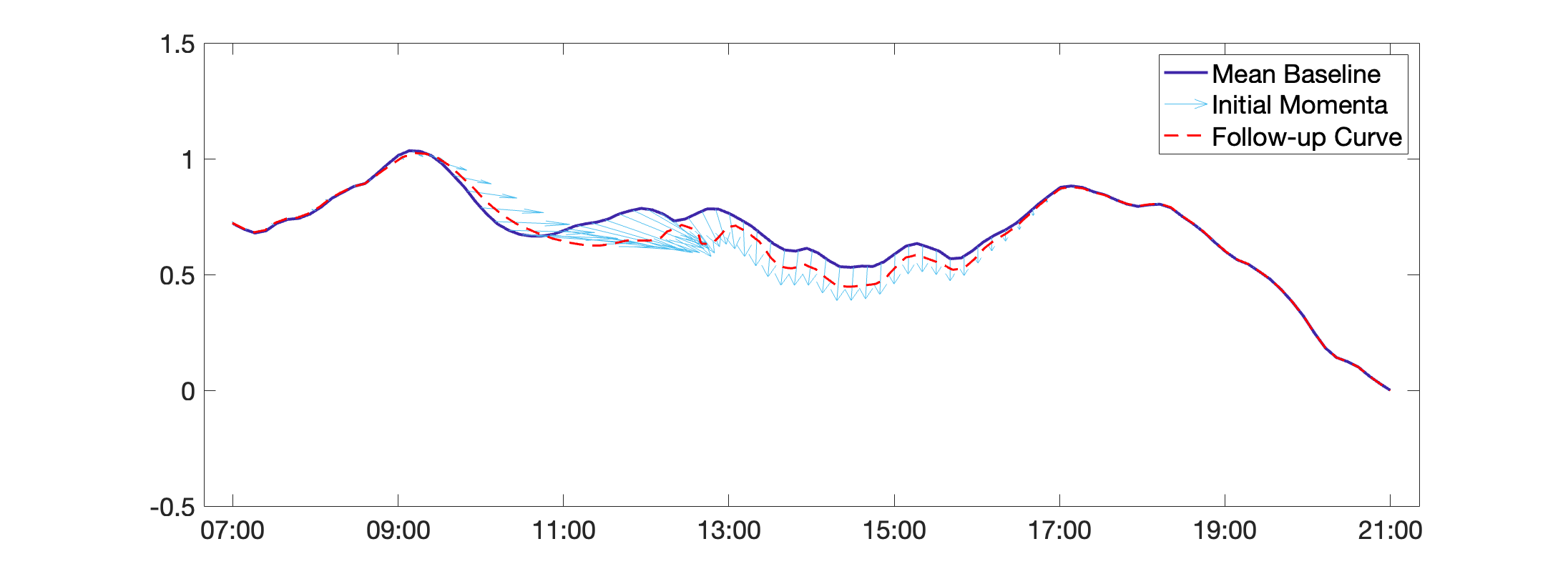}
	\caption{\small Baseline and follow-up curves of an exemplar subject}
	\label{fig:subjPC_example}
\end{figure}

We generate each subject's actual initial momenta of deformations as linear combinations of the initial momenta of the three PCs, where the coefficients in the linear combinations are simulated from independent Gaussian distributions with mean 0 and standard deviation 1, scaled by $(0.8, 1.2, 1.5)/2000$ for the coefficients of PC 1, 2, and 3, respectively, to make a distinction between the PCs. 
Each subject's follow-up PA curve is generated by deforming the baseline PA curve with the subject-specific initial momenta following the procedure described in Section \ref{sect: model and method}.
Figure \ref{fig:subjPC_example} visualizes an example subject's baseline and follow-up PA curves in which the weights are 40\%, 15\% and 45\% for PC 1, 2, and 3. The follow-up curve deformed from the baseline exhibits an apparent time shift as in PC 3, as well as a PA magnitude change throughout the day as in PC 1. The local PA magnitude change in PC 2 is less manifested due to the smaller weight.

\begin{figure}[h] 
	\centering
	\begin{minipage}[b]{0.8\textwidth}  
		\centering
		\includegraphics[width=\textwidth]{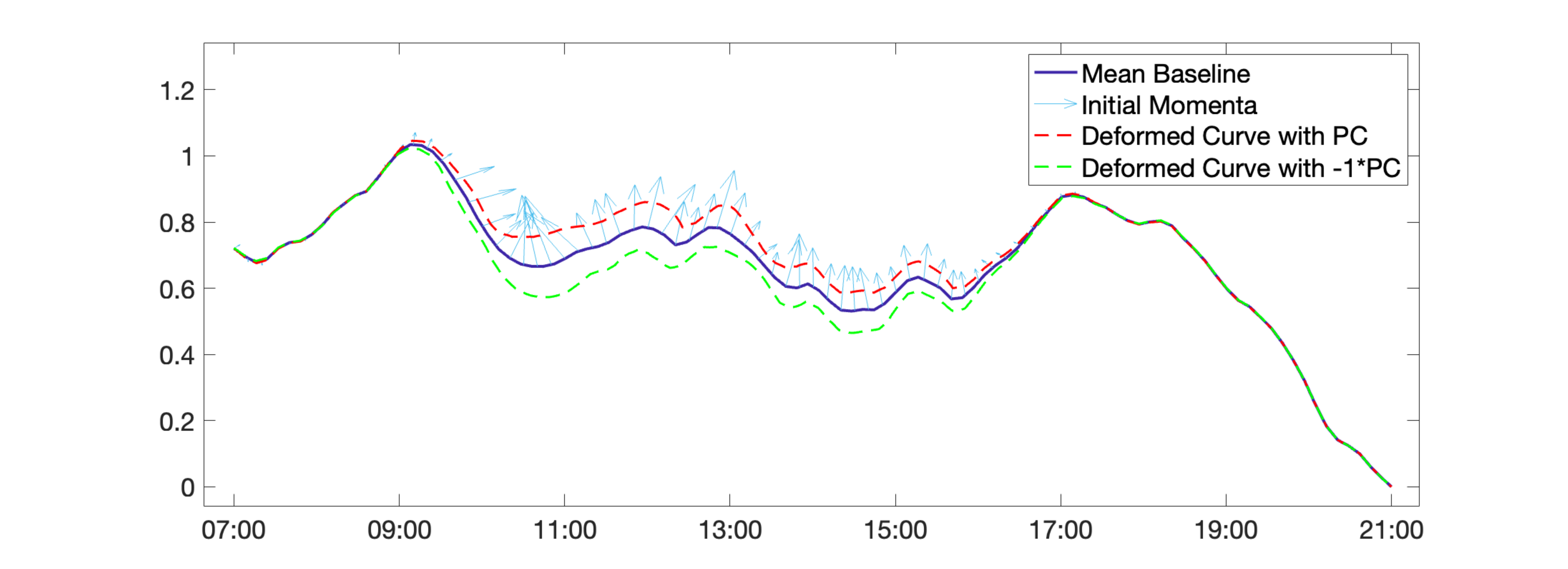}
	\end{minipage} 
	\begin{minipage}[b]{0.8\textwidth}  
		\centering
		\includegraphics[width=\textwidth]{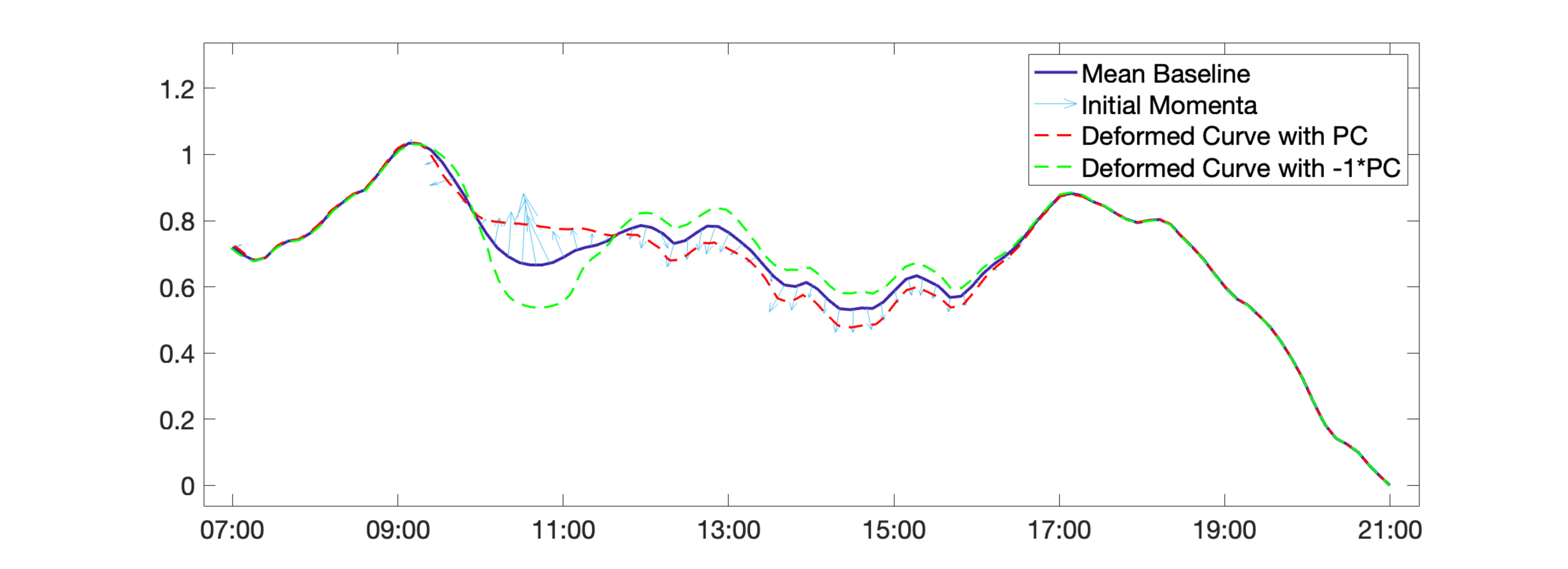}
	\end{minipage} 
	\begin{minipage}[b]{0.8\textwidth}  
		\centering
		\includegraphics[width=\textwidth]{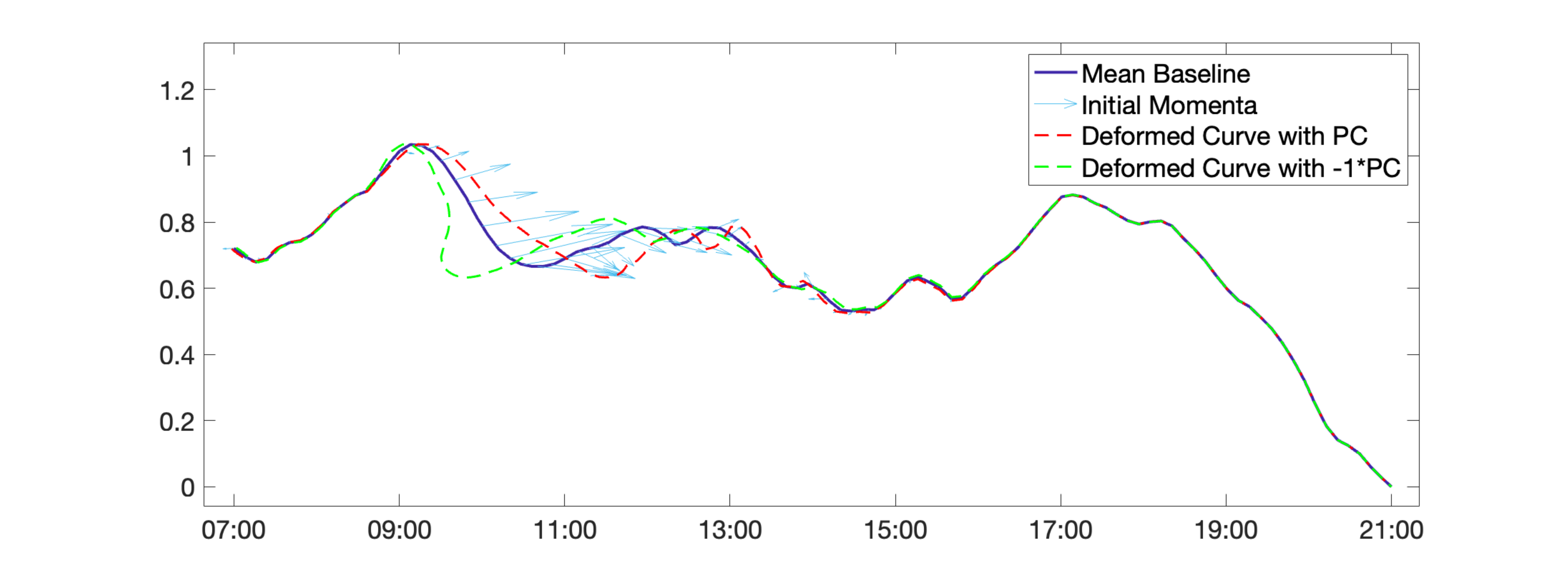}
	\end{minipage}
	\caption{From top to bottom: estimated initial momenta with proposed approach (blue arrows) of PC 1, 2, and 3 and deformations when applied to the mean baseline curve (solid navy) to deform into the follow-up curves (dashed red with PC and green with $-1 \times $ PC)} \label{fg: PC_prop}
\end{figure}

With the observed baseline and follow-up PA curves,  we estimate the subject-specific initial momenta of the deformation with the proposed method.
As a comparison, we also calculate the difference between the follow-up and baseline PA magnitudes at each minute and extract PCs only from the vertical differences in PA magnitudes. Interpolations using the R function \textit{approx()} are applied to time points where follow-up PA magnitudes are missing due to the deformation in the temporal domain.
In addition, we include in the comparison a two-step approach based on the time warping method in \cite{Wrobel2019}. 
In the first step, each subject's baseline curve is registered to the follow-up curve using the package \textit{registr2.0} (\cite{registr2}). 
Necessary formatting and standardization are applied to the curves prior to applying the package.  
The output of the registration for each subject is the set of warped time indices as a function of the original time indices in the baseline curve. 
Then functional PCs are extracted from the warped time indices using the \textit{fdapace} package.
In the second step, differences in the PA magnitude between the follow-up curves and the warped baseline curves are calculated and PCs are extracted from the differences.

\begin{figure}[h] 
	\centering
	\begin{minipage}[b]{0.8\textwidth}  
		\centering
		\includegraphics[width=\textwidth]{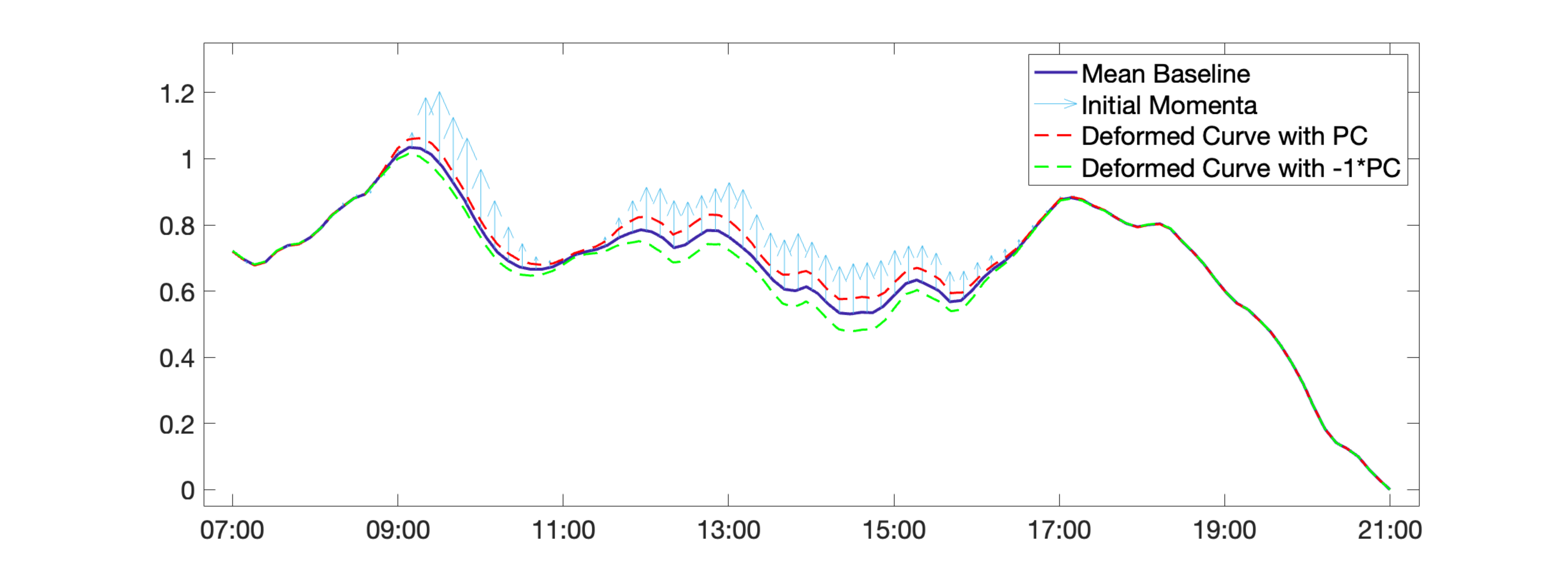}
	\end{minipage} 
	\begin{minipage}[b]{0.8\textwidth}  
		\centering
		\includegraphics[width=\textwidth]{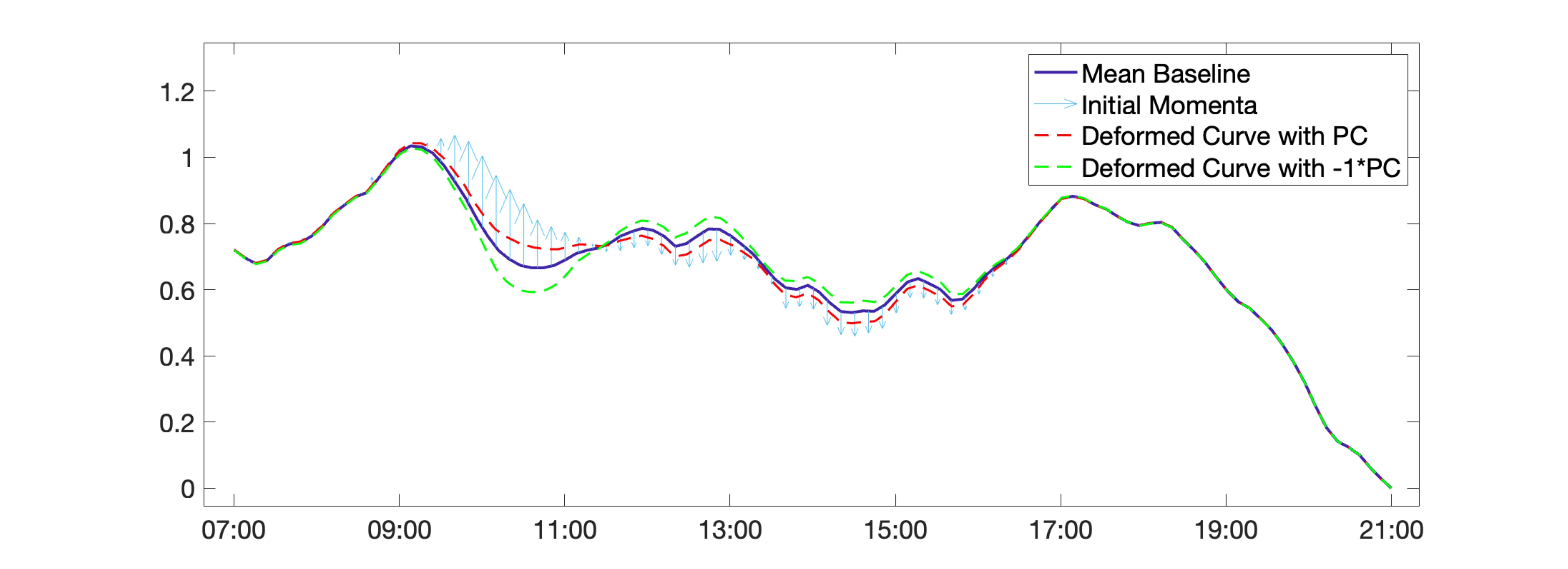}
	\end{minipage} 
	\begin{minipage}[b]{0.8\textwidth}  
		\centering
		\includegraphics[width=\textwidth]{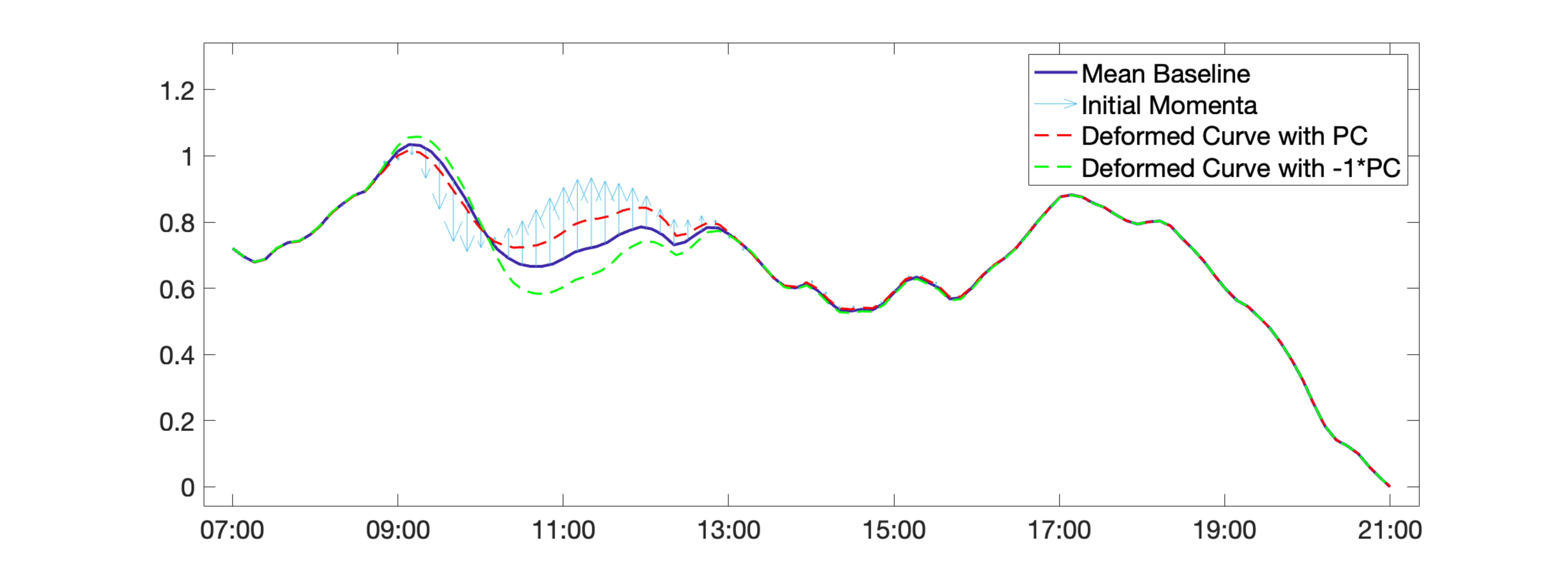}
	\end{minipage}
	\caption{From top to bottom: estimated initial momenta with PCA applied to vertical differences between baseline and follow-up PA levels (blue arrows) of PC 1, 2, and 3 and deformations when applied to the mean baseline curve (solid navy) to deform into the follow-up curves (dashed red with PC and green with $-1 \times $ PC)} \label{fg: PC_diff}
\end{figure}

Figures \ref{fg: PC_prop} and \ref{fg: PC_diff} show the estimated PCs using the proposed method and the PCs extracted from vertical differences.
The proposed method to a large extend recovers the modes of longitudinal changes in PA patterns by capturing most of the vertical changes in PA magnitudes and temporal shifts of PA patterns. 
On the other hand, analysis of only the vertical changes between the baseline and follow-up PA magnitudes is unsuccessful in revealing the major changes in PA diurnal patterns: 
the overall increase of PA in the actual PC1 is only partially recovered with missing increase patterns around 10am-12pm; 
the period of locally boosted PA in PC2 is incorrectly extended to 9am-10am.
The temporal shift in PC3 is almost completely overlooked. 
The estimated PC 3 tries to recover the shift by having a local decrease before 10am and a local increase around 10am-1pm, but was  unsuccessful in capturing the shift of PA patterns as made clear by the red dashed curves. 
Correlations between the estimated and actual PC momenta and scores are available in the Supp. Material.

\begin{figure}[h] 
	\centering
	\begin{minipage}[b]{0.8\textwidth}  
		\centering
		\includegraphics[width=\textwidth]{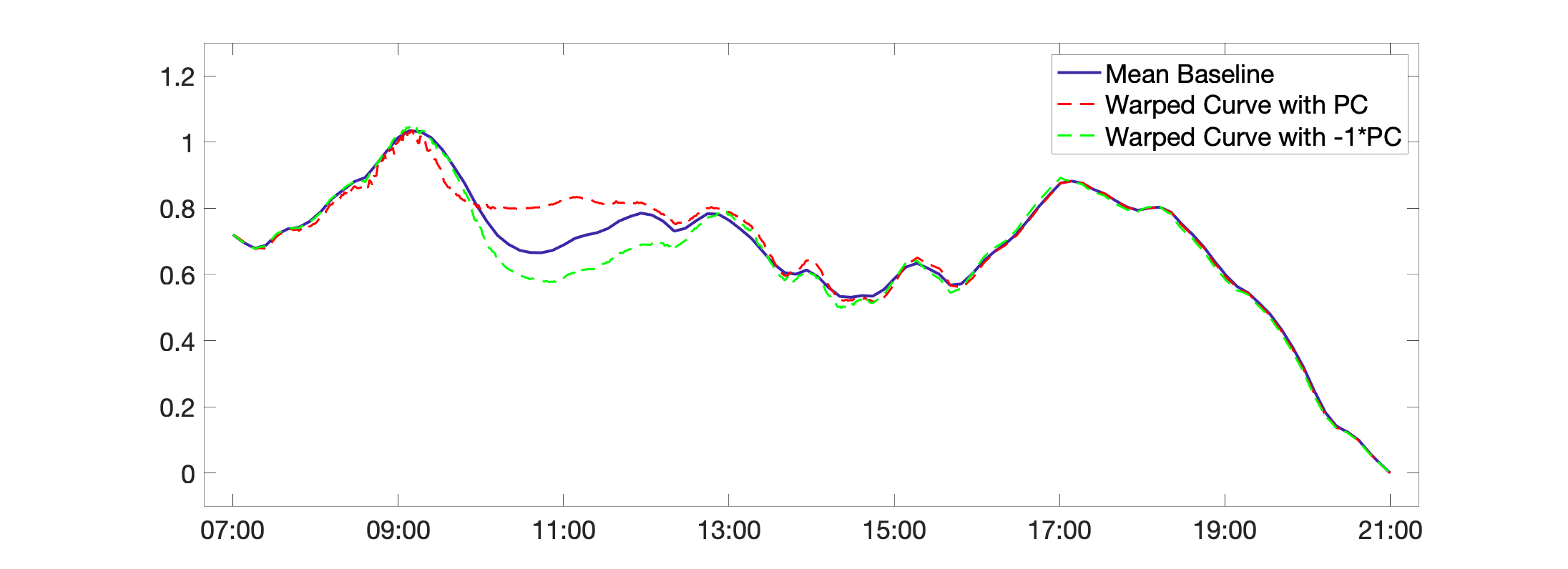}
	\end{minipage} 
	\begin{minipage}[b]{0.8\textwidth}  
		\centering
		\includegraphics[width=\textwidth]{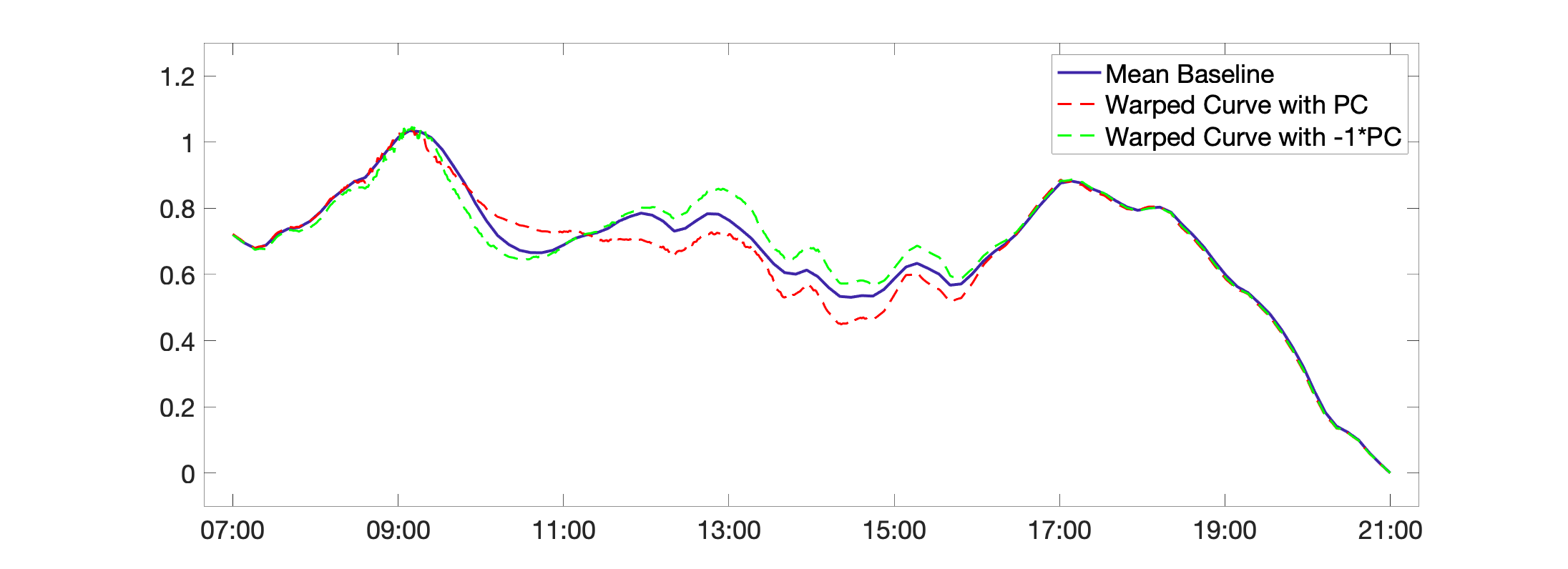}
	\end{minipage} 
	\begin{minipage}[b]{0.8\textwidth}  
		\centering
		\includegraphics[width=\textwidth]{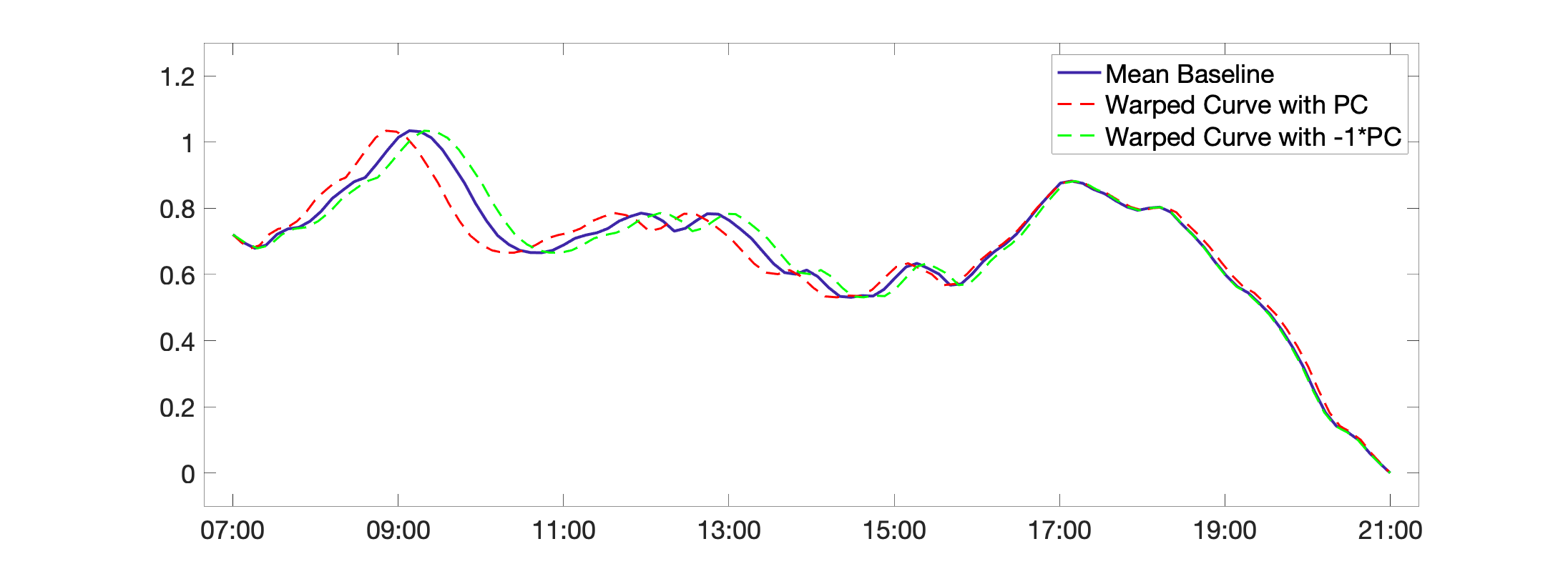}
	\end{minipage}
	\caption{Top and middle: estimated follow-up curve associated with PC 1 and 2 extracted from difference in PA magnitude between the follow-up.
		Bottom: follow-up curve associated PC 1 of time-warping of the mean baseline curve. Solid navy: baseline curve. 
		Dashed red and green: estimated follow-up curves associated with 1 and (-1) times the PCs, scaled by the standard deviation.} \label{fg: PC_wrobel}
\end{figure}

Figure \ref{fg: PC_wrobel} shows the PCs estimated from the two-step approach based on the time-warping method in \cite{Wrobel2019}. The bottom panel shows the estimated follow-up curve when applying the PC 1 extracted from the time warping functions, which explained over 97\% of the total variance in the subjects.
Note there are no momenta involved in this method and thus no arrows are plotted.
Comparing to the actual PC 3 in Figure \ref{fg: PC_actual}, the estimated PC captures most of the shift between 9am and 1pm, but tends to over-estimate the range of the shift: estimated shifts are also visible from 7am - 9am and 1pm - 3pm.
The top and middle panels show the top two PCs extracted from the vertical difference between the follow-up PA curves and the time-warped baseline curves.
Comparing to the actual PC 1 in Figure \ref{fg: PC_actual} of an overall increased/decreases change in PA magnitude between 9am and 5pm,  the estimated PC 1 is capable of recovering the change in the time window of 10am - 1pm, but fails to capture the changes outside this range.
The estimated PC 2 resembles the actual PC 2 in the local change of PA magnitude between 10am - 11am, but the estimated change is less apparent comparing to the actual PC, especially when contrasting with the change in the opposite direction in the time window of 11:30am-3pm.
There are also some mix-ups in the PC 1 and PC 2 of the PA magnitude change. 
The difference in results can be partially due to the sensitivity of the estimation of warping to the change in PA magnitude, especially to changes in small/local time windows as those in PC 2. While the an apparent overall increase/decrease in PA magnitude between the baseline and the follow-up curves can be adjusted via centering and standardization prior to applying the warping estimation, changes in local time windows are more difficult to adjust and can lead to a mix-up in the estimation of time warping and vertical change in PA magnitude. 
On the other hand, the proposed approach is capable of accounting for both changes in the temporal domain and in the vertical PA magnitude and achieve more accurate estimation of both changes.

\section{Data Analysis I: the RfH Study} \label{sect: data analysis 1}

In this section we analyze the RfH study data, which consist of PA records at baseline and month-6 for 333 overweight, postmenopausal early-stage breast cancer survivors. Each subject was randomly assigned to one of the $2\times 2$ life-style intervention and medication treatment arms. Our focus here is to study the effect of the treatments on the longitudinal changes in PA.

\subsection{Pre-processing of Data}
Subjects with both baseline and month-6 measurements available from the activity tracker are retained for the analysis. As a result, 303 out of 333 subjects remain in the dataset after cleaning for missing records. 
For each subject, the baseline and month-6 activity tracker measured PA records are sorted by dates and time of wearing.
We only use PA records of each day from 6am to midnight as most of the records out of this range can be omitted due to non-wear time when the subject was sleeping.
Within the 6am-midnight window, whenever the subject was not wearing the device, PA was recorded as zero and treated as inactive time due to sleeping, since subjects were asked to wear the activity tracker during all waking hours except for time contacting water (\cite{Xu:2019bd, Patterson:2018ck}).
Note here we use the chronological time to index the PA records instead of align subjects' PA records according to the subject-specific time of first non-zero record, as of interest here are the diurnal activity patterns as a function of chronological time and the longitudinal change in such activity patterns.

The retained PA records for each subject are then averaged over multiple days of measurements within each period and the resulting records have format $(t, \text{PA}(t))$ for each period, where $t = 1, \dots, 2160$ are minutes from 6am to midnight.
Smoothing with splines is then applied to the averaged baseline and month-6 activity PA to get the smoothed PA curves.
The smoothing reduces the variability induced by random measurement errors and fluctuations caused by the device and to increase the signal-to-noise ratio in the activity record to avoid over-fitting in the subsequent analyses.
The same smoothing parameters are used for all baseline and month-6 PA records for all subjects.
The smoothing parameter $df = 25$ is selected based on exploratory analysis of both RfH and MENU datasets. 
Finally, we center and scale the time and PA magnitudes so that they are both in the range $[-1,1]$ approximately. 
As discussed in Section \ref{sect: est momenta}, this step balances the penalties on the deformation energies of the initial momenta on both $x$ and $y$ coordinates.

In addition to PA, demographic and medical measurements were collected for each subject in both baseline and month-6 visits. 
In particular, treatment assignment indicators of medication (metformin and placebo) and life-style intervention (with and without weight-loss program) are available both separately and as a $2 \times 2$ factorial design. 
Other important variables include demographics, body mass index (BMI), activity minutes at baseline, history of diseases and medication, pathological stage, years from diagnosis, and glucose.

\subsection{Estimating Longitudinal Changes in PA} \label{sect: est momenta rfh}

The first step of the estimation is to characterize the longitudinal changes of PA in terms of subject-specific deformations from the baseline activity curves to the month-6 curves.
As discussed in Section \ref{sect: model and method}, each subject's deformation from the baseline curve to the month-6 curve is modeled as a diffeomorphism governed by a vector field, and the vector field is assumed to follow the RKHS representation \eqref{eq: rkhs}. To estimate the deformations, it suffices to estimate the initial momenta at the control points.

Here we use the function \textit{fsmatch\_tan()} in the Matlab package \textit{fshapesTK} (\cite{Charlier:2015fha}) to estimate the subject-specific initial momenta in deforming the baseline to the month-6 PA curve. 
Recall the deformation at each middle step $\tau \in (0,1)$ follows the differential equation $ \partial \psi_{\boldsymbol{v}}(\tau, \cdot)  /\partial \tau = v_\tau \circ \psi_{\boldsymbol{v}}(\tau, \cdot)$. 
In estimating the momenta, we discretize $\tau$ into finite values $\{\tau_k: k = 1, \dots, N_{\text{steps}}\}$ and estimate each $v_{\tau_k}$ in each step.
Upon examining the discretization we find $N_{\text{steps}} = 11$ steps lead to consistently satisfactory approximation of the deformation.
Figure \ref{fig:deform_example} shows estimated deformation of an example subject including the deformed curve at $\tau = 4/11$.
For data of much higher dimensionality, exploratory analysis can be conducted to find the appropriate number of steps in the discretization.
It takes  about 2 minutes to estimate the deformation from the baseline to the follow-up PA curve for each subject.  High computing cluster is used to run the estimating procedure for all subjects in parallel.
Details on function parameters in the estimation are available in the Supp. Material.


\subsection{fPCA Estimation}

The estimated initial momenta of the deformations are concatenated into vectors and the principal components (PCs) of the vectors are extracted from the mean-subtracted momenta using the function \textit{FPCA()} of R package \textit{fdapace} (\cite{fdapace}). 
The top 30 explain about 70\% of the variance in the data. 
In what follows, we will focus on the analysis involving the top 10 PCs that explain about 40\% of the variance.
Other PCs that explain less variability in the data and results are deferred to Supp. Material.

\subsection{Intervention and Covariate Associations With PA Change}

We are interested here in the effects of the covariates, the interventions on activity in particular, on different modes of longitudinal changes in PA. 
To this end we conduct an analysis on each of the top $10$ PCs. 
In each separate analysis, the projections (scores) of subject-specific deformations on each PC is the outcome of interest, and life-style intervention and relevant demographic and medical variables are included as covariates.
Table ``RfH Summary Statistics'' in the Supp. Material lists summary statistics of the interventions and other variables.
Variables of  with more than 5\% values missing are discarded for further analysis. 
Medical conditions with less than 5\% missing are imputed with ``no'' for the missing values.

Lasso regularization with 10-fold cross-validation for the choice of penalty parameter is used to select variables from the covariates.
Several variables are discovered as significantly associated with modes of change in PA as characterized by certain PCs. 
In what follows, we focus on results of PC 1 and 3, which are found to be associated with lifestyle intervention. 
Significant effects of covariates on the corresponding PC projection coefficients are listed, and initial momenta of PCs in question are visualized along with the estimated month-6 curve resulting from deforming the mean baseline curve of all subjects according to the diffeomorphism governed by the initial momenta of the PC.
For the purpose of demonstration, we also visualize the month-6 curve resulting from deforming the baseline with the estimated initial momenta of the PC multiplied by $-1$ to illustrate the opposite direction of the PA change if a subject has negative projection coefficient on the PC.
Note each PC can be scaled by a positive or negative constant and still be a well-defined PC as long as the PC scores change are scaled accordingly. 
For simplicity in interpreting the results, we scale the PCs and corresponding projections/scores so that subjects' projections of each PC have standard deviation equal to one.

\subsubsection{PC 1}

\begin{figure}[H]
	\centering
	\includegraphics[width=0.9\linewidth]{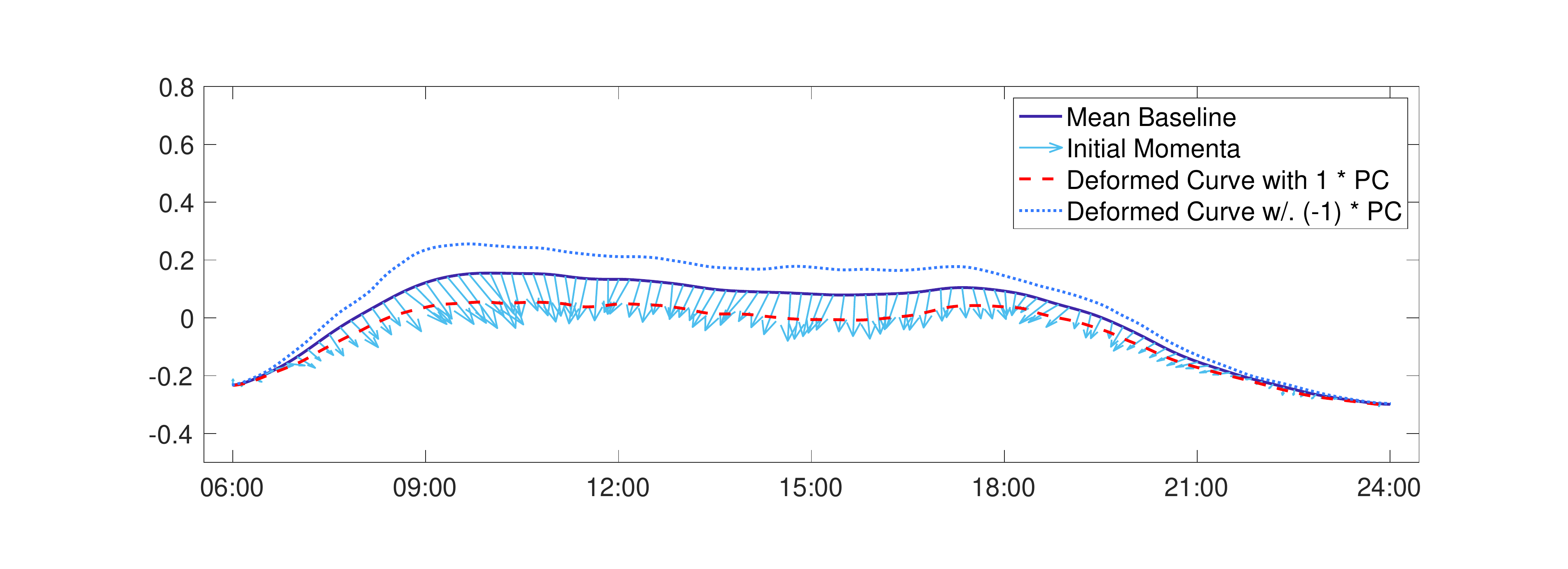}
	\caption{\small RfH study: estimated deformations of PC 1. Baseline (navy), estimated initial momenta (arrows on top of the baseline curve) and estimated month-6 curves by applying the deformations corresponding to $+1$ (red) and $-1$ (dashed blue) times the initial momenta of the PC}
	\label{fig: PC1}
\end{figure}

Figure \ref{fig: PC1} visualizes the first PC.
As PC 1 explains most variability in the data, it is not surprising that the mode of change in PA characterized by PC 1 is an overall decrease of activity level from the baseline to month 6 (or overall increase if consider $-1 \times $ PC 1).
In particular, activity levels in the morning from around 9am to 5pm change most significantly as illustrated by the downward pointing arrows.

In modeling effects on PC 1 while adjusting for other covariates, life-style intervention is selected to be included in the model with Lasso regularization. The Lasso penalty parameter is chosen based on a 10-fold cross validation.
Table \ref{tab: rfh_PC1} shows the result of a re-run of a separate un-regularized regression analysis using only variables selected by Lasso regularization. 
Life-style intervention is significant with a negative effect $-0.265$, and hence a positive effect on $-1 \times $ PC 1, with $p$-value $0.015$, indicating the intervention potentially has a positive effect on promoting daily activity.

\begin{table}[H]
	\centering
	\small
	\begin{tabular}{|l|l|l|}
		\hline
		\textbf{Covariate} & \textbf{Coef Estimate (SE)} & \textbf{p-value} \\ \hline
		(Intercept) & -3.617 (0.920) & \textless{}0.001 *** \\ \hline
		LifeStyle Intervention & -0.265 (0.108) & 0.015 * \\ \hline
		AGE & 0.027 (0.027) & 0.312 \\ \hline
		Age at diagnosis & 0.005 (0.026) & 0.843 \\ \hline
		SED & -0.001 (0.001) & 0.504 \\ \hline
		LPA100 & 0.004 (0.001) & \textless{}0.001 *** \\ \hline
		MVPA1952 & 0.008 (0.003) & 0.012 * \\ \hline
		Leptin & 0.005 (0.002) & 0.036 * \\ \hline
		Baseline Weight & 0.005 (0.004) & 0.227 \\ \hline
		Insomnia Yes & -0.110 (0.118) & 0.352 \\ \hline
		HighCholesterol Yes & 0.198 (0.108) & 0.066 . \\ \hline
		Diabetes Yes & -1.012 (0.355) & 0.005 ** \\ \hline
		Aspirin Yes & -0.373 (0.127) & 0.003 ** \\ \hline
		Education Graduated from  high school or G.E.D. & -0.408 (0.222) & 0.067 . \\ \hline
		Marital Living with a partner in a marriage like relationship & -0.408 (0.282) & 0.149 \\ \hline
		Employment Disabled and/or retired because of health & 0.449 (0.220) & 0.043 * \\ \hline
	\end{tabular}
	\caption{RfH study: coefficient estimates (with standard errors) and $p$-values of covariates' effects on PC 1 projection scores.} \label{tab: rfh_PC1}
\end{table}

Table \ref{tab: rfh_PC1} also lists estimated effects of Lasso-selected covariates on overall decrease in PA as characterized by a positive projection score on PC 1.
Cut-point based average minutes categories at baseline: LPA100 (Light Physical Activity with CPM, count per minute of the VM, in the range of $100 - 1951$), and MVPA1952 (Moderate Vigorous Physical Activity with CPM $> 1952$) are significantly correlated with PC 1. For details on above cut-point based categories, see, for example, \cite{Amagasa:2018hb}.
Signs of the effects associated with the above categories (positive for both LPA100 and MVPA1952) are reasonable, as those who are less active and more sedentary are expected to have more space of improvement and thus exhibit an increase in PA during the study.
Additional diagnostics plots are available in the supplementary material.

Subjects with diabetes tend to exhibit increased activity. Given the relatively small number of subjects with diabetes (7 out of 303), however, the effect of diabetes needs to be validated in a more comprehensive analysis of a more balanced dataset.
Taking Aspirin (such as Anacin, Bufferin, Bayer, Excedrin) regularly more than three times per week at baseline visit is also associated with increased activity.
Employment status of disabled and/or retired because of health is associated with decreased PA comparing to other employment status groups.
Lastly, the randomized assignment of medication (metformin or placebo) is not selected by Lasso.

\subsubsection{PC 3}

\begin{figure}[H]
	\centering
	\includegraphics[width=0.85\linewidth]{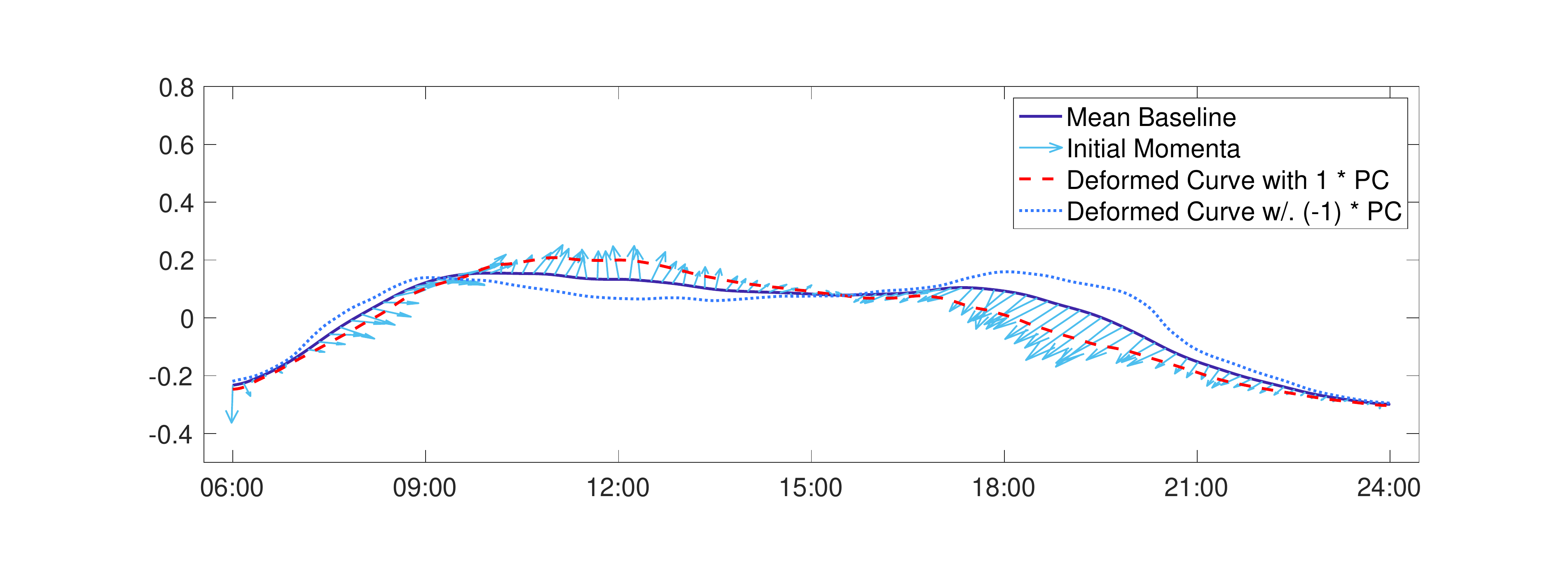}
	\caption{\small RfH study: estimated deformations of PC 3}
	\label{fig: PC3}
\end{figure}

Figure \ref{fig: PC3} visualizes the estimated PC 3. 
The mode of variation in PA associated with PC 3 is marked by a mild increase of PA level at mid-day (10am - 2pm) and a decrease in the evening (6 - 9pm).
In addition, there is a possible shift of PA pattern to early hours at around 6-8pm and a less manifested shift to later hours in the morning between 6-9am.

\begin{table}[h]
	\centering
	\small
	\begin{tabular}{|l|l|l|}
		\hline
		\textbf{Covariate} & \textbf{Coef Estimate (SE)} & \textbf{p-value} \\ \hline
		(Intercept) & -1.076 (0.642) & 0.095 . \\ \hline
		LifeStyle Intervention & -0.195 (0.112) & 0.082 . \\ \hline
		Wear time & 0.002 (0.001) & 0.033 * \\ \hline
		Smoke Status Former & -0.365 (0.112) & 0.001 ** \\ \hline
		Employment Disabled and/or retired because of health & -0.460 (0.232) & 0.048 * \\ \hline
	\end{tabular}\caption{RfH study: coefficient estimates (with standard errors) and $p$-values of covariates' effects on PC 3 projection scores.} \label{tab: rfh_PC3}
\end{table}

Table \ref{tab: rfh_PC3} summarizes estimated effects of Lasso-selected covariates in the re-run of regression for PC 3.
Life style intervention has a significantly negative effect, indicating subjects in the intervention group are more likely to exhibit a change in PA in the direction of $-1 \times $ PC 3 (as illustrated by the blue dashed line in Figure \ref{fig: PC3}), with lower PA levels between 9am-2pm but higher PA after 6pm and a shift of PA pattern to later hours.
Again randomized assignment of medication (metformin or placebo) is not selected by Lasso as a relevant predictor.
Moreover, being a former smoker ($n=135$) is associated with aforementioned mode of change in PA patterns, comparing to those who never smoke ($n=164$) or current smokers ($n=4$). 
The same phenomenon is observed for employment status of disabled and/or retired because of health.
Wear time is positively associated with PC 3, indicating those who wore the device longer tend to exhibit increased PA levels in the middle of the day but decreased PA in the evening.


\section{Data Analysis II: the MENU Study} \label{sect: data analysis 2}

The MENU study data consist of 245 overweight non-diabetic women to study the effect of randomly assigned diet interventions on changes in health outcomes in 12 months. Of particular interest is the effect of the diet interventions on weight loss. 
In line with this aim, for the analysis of the MENU data, we focus on the role of longitudinal changes in PA on weight loss characterized by changes in BMI.


A total of
$177$ subjects with both baseline and month-12 PA records available were retained for the analysis after processing.
Upon observing the raw PA data, only PA records within the range of 6am to midnight are kept for the analysis as data out of this range were mostly inactive records.
Same pre-processing procedures are applied to the MENU study data as for the RfH data. 
The same estimation pipeline for the subject-specific deformations between the baseline and month-12 PA curves used for the RfH data is applied to the MENU data. 
fPCA is applied to the estimated initial momenta for each subject's deformation from the baseline PA curve to the month-12 curve with R package \textit{fdapace} (\cite{fdapace}). 
Top 30 PCs explain over 86\% of the variance.


%



The outcome of interest here is the change in BMI, that is, month-12 minus baseline BMI, as an indicator for weight loss during the 12-month study period.
We examine models in which change in BMI is the outcome and subjects' projection coefficients on the PCs are the main covariates of interest. 
Other important covariates including diet interventions, demographics, smoking, medical history, and baseline activity metrics defined by cut-points are adjusted in the model.
Table ``MENU Summary Statistics'' in the Supp. Material lists summary statistics of the interventions and other variables.

\begin{figure}[h]
	\centering
	\includegraphics[width=0.9\linewidth]{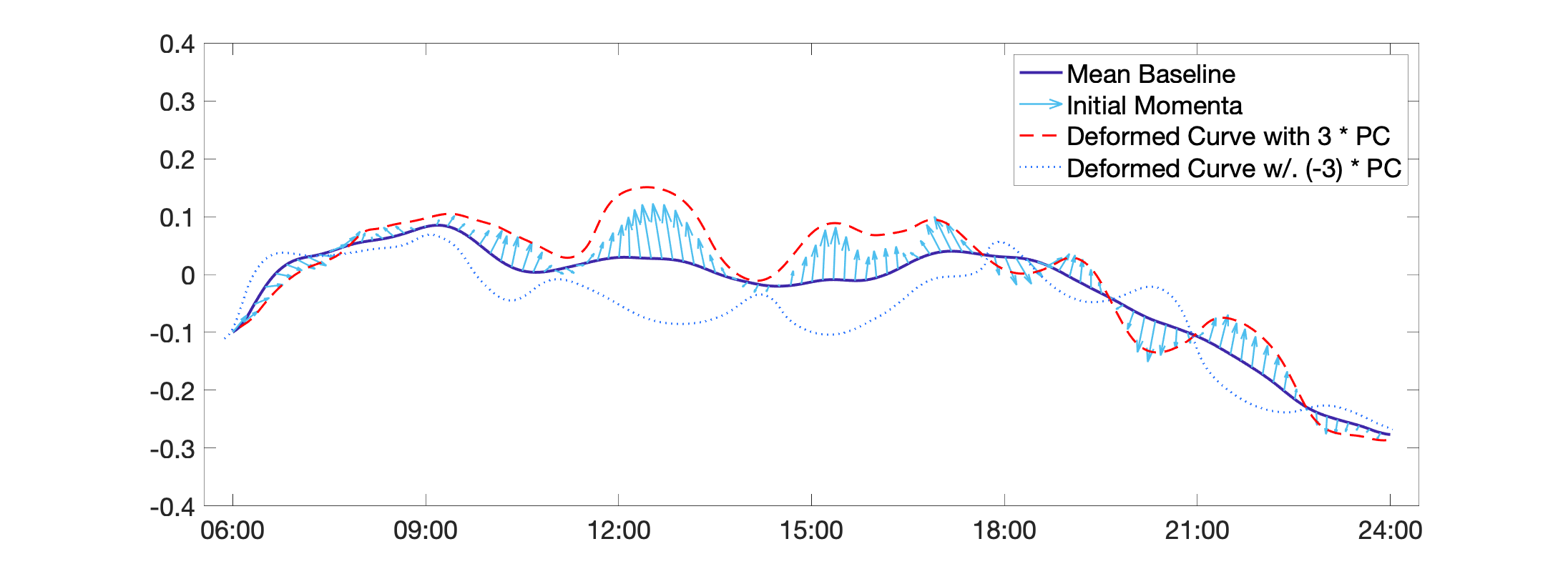}
	\caption{\small MENU Study: estimated deformations of PC 10.}
	\label{fig: MENU_PC10}
\end{figure}

The analysis consists of two steps.
First, we fit a regression with Lasso regularization to select variables including the PCs and other covariates that are associated with the outcome.
Second, we fit a separate regression including only Lasso-selected variables as well as race, marital status, and diet intervention, which are also of interest.
Table \ref{tab: MENU reg} lists the result of regression in the second step of analysis.
PCs 1, 10, 20, 25 are discovered to be significantly associated with change in BMI.
Model diagnostics plots are also available in the supplementary material.
Figures \ref{fig: MENU_PC1} and \ref{fig: MENU_PC10} illustrate the estimated deformations associated with PCs 1 and 10. 
Plots of PCs 20 and 25 are deferred to the Supp. Material as they explain a smaller portion of the variance.

\begin{table}[h]
	\centering
	\small
	\begin{tabular}{|l|l|l|}
		\hline
		\textbf{Covariate} & \textbf{Coef Estimate (SE)} & \textbf{p-value} \\ \hline
		Intercept & -0.060 (0.825) & 0.942 \\ \hline
		Diet Lower Fat & -1.123 (0.479) & 0.020 * \\ \hline
		Diet Walnut-Rich & -0.772 (0.498) & 0.123 \\ \hline
		Black & 0.608 (0.842) & 0.471 \\ \hline
		Asian & -1.043 (1.478) & 0.481 \\ \hline
		Pacific Islander & -2.112 (1.820) & 0.248 \\ \hline
		Native American & -2.380 (2.610) & 0.363 \\ \hline
		Mixed Race & -0.978 (1.470) & 0.507 \\ \hline
		Other Race & 3.888 (2.542) & 0.128 \\ \hline
		Single-never married & -1.054 (0.578) & 0.070 . \\ \hline
		Widowed & -2.779 (1.301) & 0.034 * \\ \hline
		Divorced & -0.187 (0.557) & 0.738 \\ \hline
		Separated & -0.977 (2.576) & 0.705 \\ \hline
		High blood pressure & 0.229 (0.612) & 0.709 \\ \hline
		High blood pressure and High cholesterol & 0.457 (0.620) & 0.462 \\ \hline
		High blood pressure and other & 0.845 (1.806) & 0.640 \\ \hline
		High cholesterol & 1.841 (0.629) & 0.004 ** \\ \hline
		Other & 0.943 (1.178) & 0.424 \\ \hline
		Pain SF-36 subscale & -0.018 (0.009) & 0.051 . \\ \hline
		Follicle stimulating hormone (FSH) & -0.018 (0.006) & 0.003 ** \\ \hline
		PC1 & -0.507 (0.195) & 0.010 * \\ \hline
		PC10 & -0.428 (0.192) & 0.027 * \\ \hline
		PC20 & -0.404 (0.200) & 0.045 * \\ \hline
		PC25 & -0.391 (0.199) & 0.051 . \\ \hline
	\end{tabular}
	\caption{MENU study: coefficient estimates (with standard errors) and $p$-values on $\text{BMI}_\text{month 12} - \text{BMI}_\text{baseline}$}
	\label{tab: MENU reg}
\end{table}

\begin{figure}[h]
	\centering
	\includegraphics[width=0.9\linewidth]{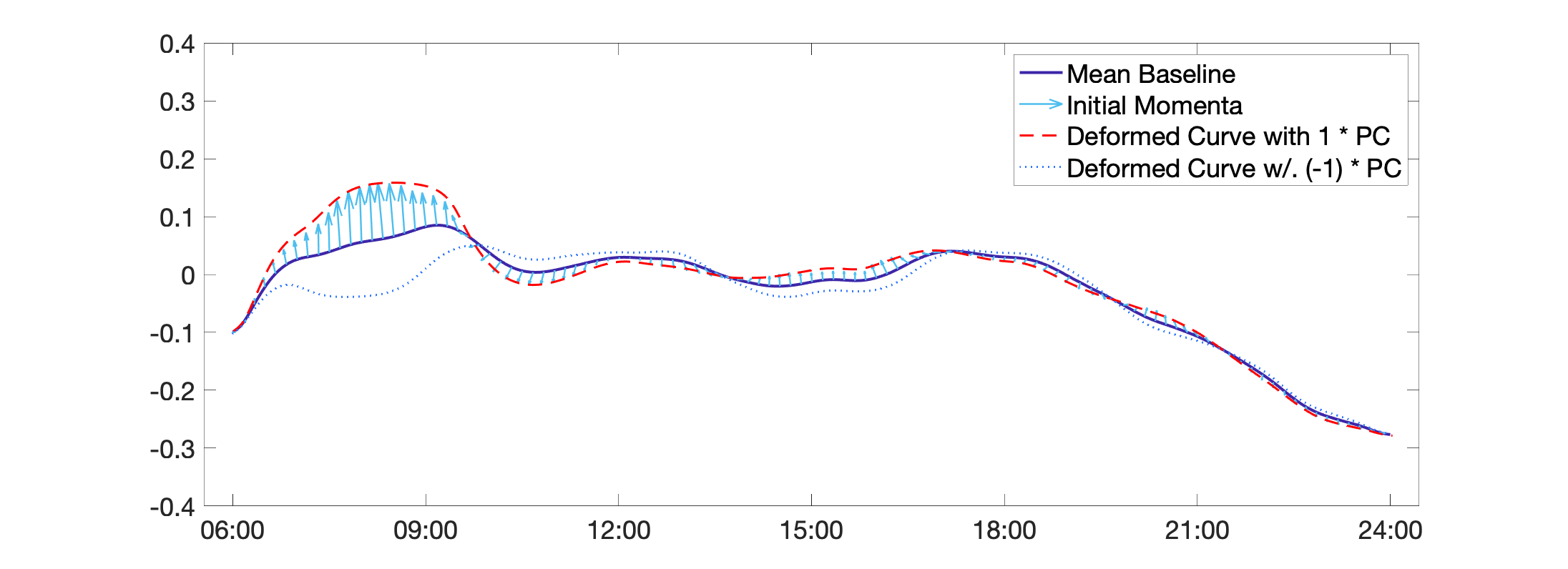}
	\caption{\small MENU Study: estimated deformations of PC 1.}
	\label{fig: MENU_PC1}
\end{figure}

Interestingly, PC 1 of MENU is not an overall increase/decrease of PA throughout the day as discovered in the RfH study.
The red curve in Figure \ref{fig: MENU_PC1} shows that the mode of variation characterized by PC 1 is a local increase of activity in the morning between 6 - 10am, as shown by the upward pointing blue arrows. 
The deformed curve from the baseline following the PC's initial momenta is illustrated by the solid red curve (PC multiplied by -1 as illustrated by the blue dashed curve).
The effect of PC 1 on the change in BMI is negative (-0.507 with $p$-value 0.01), indicating an increase of morning PA from baseline to month-12 is associated with a decreased BMI.

Figure \ref{fig: MENU_PC10} shows the mode of change in PA associated with PC 10, which exhibits a re-distribution of PA. 
Activity levels are increased significantly around 12 - 2pm, 3pm - 5:30pm, and 9 - 11pm, while decreased slightly around 6 - 7pm and 8 - 9pm.
There is also a slight shift of PA towards later hours in the early morning shown by the arrows pointing to the right at time window 6 - 8am.
The effect of PC 10 on the change in BMI is also negative (-0.428 with $p$-value 0.027). This result is not surprising, as PC 10 exhibits mostly increased PA in different time windows of a day, and thus is expected to be associated with decreased BMI.

Diet interventions are significantly associated with change in BMI. Comparing to the baseline of lower carbohydrate and higher monounsaturated fat diet, 
the lower fat and higher carbohydrate diet has a negative effect on change in BMI ($p$-value 0.02). 
The Walnut-rich diet also has a mildly negative ($p$-value $0.123$) effect.
In addition, high cholesterol is significantly associated with positive change in BMI. 
Larger pain subscale indicates less pain and is associated with decreased BMI.
Higher follicle stimulating hormone (FSH) level is associated with decreased BMI.
Comparing to the baseline of married, marital status of single or widowed are both associated with decreased BMI.

\subsection{Alternative Approach with Functional Regression} \label{sect: func_reg}

In addition to the functional principal component analysis, we also examine the effect of change in PA diurnal patterns on changes in BMI with the functional regression approach.

The regression coefficients are estimated with methods proposed by \cite{Goldsmith2012}. The estimated initial momenta of all subjects in the $x$ (temporal) and $y$ (vertical) coordinates are treated as functional predictors of interest, and the change in BMI is the one-dimensional outcome. 
For each subject, the initial momenta of deformation between the baseline and month-12 PA curves is a vector of $1080$ elements in both the $x$ and $y$ coordinates, indicating the directions and magnitudes of changes in PA at each of the minutes between 6am and midnight. 
Formally, the model is
$\text{bmi}_\text{month 12} - \text{bmi}_\text{baseline} = \text{intercept} + \alpha Z + \int_t \beta_x(t) m_x(t) dt + \int_t \beta_y(t) m_y(t) dt + \epsilon$, 
where $(m_x(t), m_y(t))$ denote the initial momenta of deformation in PA at time $t$ for a generic subject, 
$(\beta_x(\cdot), \beta_y(\cdot))$ denotes the corresponding functional coefficients/effects, 
$Z$ and $\alpha$ denote time-invariant covariates and corresponding effects,
and $\epsilon$ denotes the error.

\begin{figure}[h]
	\centering
	\includegraphics[width=0.8\linewidth]{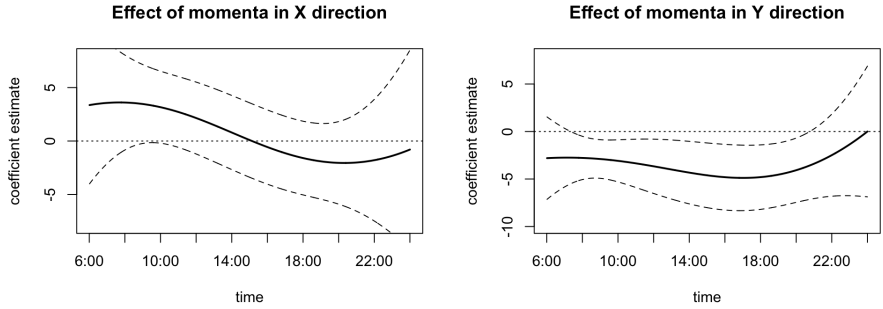}
	\caption{\small Estimates of functional coefficients adjusting for covariates.}
	\label{fig: funreg_cov}
\end{figure}

%
Model parameters are estimated with R package \textit{funreg} by \cite{funreg}.
We examine the effects while adjusting for important covariates and potential confounders selected by the Lasso model in the previous section. 
Figure \ref{fig: funreg_cov} shows the estimates of $\beta_x(\cdot)$ (left panel) and $ \beta_y(\cdot)$ (right panel) with 95\% confidence bounds. 
The effect of temporal shift is not significant.
The effect of vertical change in PA magnitudes are significantly negative between 8am - 9pm, indicating increased PA in this time window is associated with decreased BMI.
For detailed estimated effects of all covariates in the model, see the Supp. Material.
Comparing to results of fPCA, the functional regression result indicates an overall benefit of enhanced PA levels throughout the day (8am - 9pm), without providing information on detailed PA patterns.

\section{Concluding Remarks} \label{sect: conclusion}

In this paper we propose a new framework for longitudinal changes in physical activity (PA) and relations between longitudinal changes in PA and health outcomes as well as interventions. 
The model and method proposed here are based on a Riemann manifold representation of spline-smoothed PA records as functions of chronological time of a day. 
The longitudinal changes in PA during a multi-month study period are modeled as deformations between PA curves (1D Riemann manifolds) measured in different visiting periods of the study.
The deformations are modeled via diffeomorphisms governed by elements in a reproducing kernel Hilbert space that satisfy minimal-energy constraints.
Subject to the constraint, the deformation is determined by the key quantity ``initial momenta'', which are vector fields that represent the initial directions and magnitudes to ``drag'' each point on the baseline curve towards the target curve.

The variability in longitudinal changes of PA within a cohort of subjects is modeled through the variability of the deformations. 
Specifically, we adopt the functional principal component analysis (fPCA) to model the variability in subject-specific initial momenta.
We focus on top principal components (PC) that are capable of explaining most of the variability in the cohort and examine the corresponding modes of variation in longitudinal changes in PA.
In studying the longitudinal changes in PA of the subjects, of interest are the projection coefficients on the PCs for each subject, which characterize the composition of difference modes of changes in PA for each subject.

In modeling the relations between changes in PA and health outcomes/interventions, subjects' projection coefficients on PCs are used as the proxy for longitudinal changes in PA and are linked to the outcomes/interventions.
We apply the proposed model and method to data from two clinical trials: RfH and MENU. 
For the RfH we use a linear model to examine the effect of the lifestyle intervention on projections on the top PCs while adjusting for relevant covariates selected by Lasso regularization. 
PCs characterizing different modes of changes in PA diurnal patterns, including an overall boosted PA throughout the day and change of PA levels in mid-day and evening, are discovered to be significantly associated with the intervention. 

For the MENU study we use a regression model to study the effect of different modes of changes in PA, as characterized by the top PCs, on the change in BMI while adjusting for relevant covariates selected by Lasso regularization.
A significant boost of morning PA levels is found to be significantly associated with decrease in BMI.
A re-distribution of PA with increased PA around noon and 3pm is also found to be beneficial for weight loss.
Moreover, findings of PA diurnal patterns are different in the MENU and RfH studies, indicating the importance of understanding
detailed characteristics of PA in different study cohorts.

Models considered here can be readily generalized to study a variety of problems involving longitudinal changes in PA. 
For example, the effect of PA can be examined in a mediation analysis where the PC projections are potential mediators on the pathway from diet/lifestyle intervention to health outcomes. The PCs can also be studied in machine learning models to predict health outcomes.

Comparing to existing methods,
the proposed model and method reveal important information from the minute-level activity tracker measured PA data and
enable discovery of previously overlooked modes of changes in PA diurnal patterns. 
In particular, boost of activity in certain time windows and/or shifts of active hours to certain periods are found to be associated with lifestyle intervention and can be more effective in facilitating weight loss.
Subjects with certain characteristics may have higher tendencies to experience changes in PA in certain modes.
Such information is valuable in advising and treating patients, and in designing individualized intervention programs and guidelines according to subject's personal medical history and conditions.

We would also like to emphasize the difference in the focus between the proposed approach and the time warping/registration methods including \cite{Wrobel2019}.
In the latter it is assumed there exists an intrinsic time that is possibly different from the chronological time that indexes the diurnal patterns in physical activity. 
The time-warping approach aims to align the curves, usually repeatedly measured in the same period, based on the unobserved intrinsic time to reduce variability and to clarify underlying patterns.
The focus of our study is to understand longitudinal changes in PA diurnal patterns as indexed by the chronological time. 
In particular, shifts in PA patterns including active hours are under examination as a main quantity of interest.

The comparison between the two-step approach of separating temporal/phase and vertical/magnitude changes and the unified approach of modeling both changes simultaneously can also lead to many intriguing discussions. 
One important discussion is around the identifiability of the models. 
In a two-step approach, different choices of warping functions used in the first step of phase registration will lead to heterogeneity in the vertical change of PA magnitude in the second step and hence difference in the results. Essentially, there is an identifiability issue when separating the variability in the temporal domain and in the magnitudes.
A relevant discussion can be found in \cite{marron2015functional} and references therein, where the authors discuss about the information gained from the unified approach comparing to the two-step approach.
Some other interesting questions around this comparison include whether the two steps should be interchangeable in the two-step approach, and the choice of warping based on parametric models and non-parametric models. Due to the page limit these questions are better suited for a thorough examination in future studies.

One of the advantages of directly modeling the change in PA via the diffeomorphism is that subjects' heterogeneity in each period's PA can be partially eliminated in estimating the subject-specific deformations between the PA curves in different periods.
By directly modeling the change in PA, we are able to focus on the analysis of variability in the change with less interference from the heterogeneity in each individual period. 

Potential sleep/wake up time change is accounted for in the model of the diffeomorphisms.
For example, if there is a change for a subject in the sleep and wake up time between the baseline and the follow-up period, it will be reflected by a temporal shift of the earliest active time in the PA curve. 
In fact, the proposed model is versatile in characterizing not only changes in the active hours where PA happened, but also changes in the transition between non-active and active hours. 
If seasonality effect is of concern, adjustments can be achieved by a thorough exploratory analysis and pre-processing of the data. Adjustments can also be applied to the statistical inference involving the estimated diffeomorphisms.

In future studies, we will explore details in modeling PA records collected from more than two visits. 
In addition, we are interested in further methodology development for statistical inference on changes in PA characterized by deformations of PA curves. 
For example, functional regression approaches \cite{Reiss2007, Crainiceanu2009,Goldsmith2012} can be utilized to study effects of covariates on longitudinal changes in PA and/or effect of PA changes on health outcomes. 
In this paper we provide some first results of the MENU study with functional regression models in which the longitudinal change in PA represented by the deformation momenta is a functional covariate of interest. The deformation momenta on the $x$ and $y$ coordinates are modeled as separate covariates. In future studies, we aim to explore modeling the deformation momenta as two-dimensional random fields and study theoretical properties of the model. 

There can be multiple directions following this line of research based on the modeling of longitudinal changes in PA with diffeomorphisms. One of such directions is clustering of subjects in a study based on the modes of changes in PA patterns, where the clustering can be conducted based on similarity metrics constructed for the diffeomorphisms.
Another possible direction is to study potential causal relationships between changes in PA and health outcomes and interventions including potential mediation effects. Alternative time warping approaches to align PA curves incorporating changes in the PA magnitudes can also be developed based on this model.

Finally, we will organize the algorithm pipeline of the proposed procedure, including pre-processing and smoothing of raw accelerometer-measured PA records, estimation of deformation between PA curves in different periods, and estimation of correlative model of PA and health outcomes/interventions.
We plan to publish source code and example vignettes of  software developed on open-source platforms such as GitHub to share with global researchers so that the proposed approach can be readily applied and results can be replicated.


%
%
%
%
%
%
%

\bibliographystyle{apalike}
\bibliography{ref_new}

\end{document}